# Interlayer couplings in homobilayer structures of $M$Si$_2$X$_4$ ($M$ = Mo/W, $X$ = N/P/As)

Qian Wang[1], Na Zhang[1], Hongyi Yu[1,2*]

[1] Guangdong Provincial Key Laboratory of Quantum Metrology and Sensing & School of Physics and Astronomy, Sun Yat-Sen University (Zhuhai Campus), Zhuhai 519082, China
[2] State Key Laboratory of Optoelectronic Materials and Technologies, Sun Yat-Sen University (Guangzhou Campus), Guangzhou 510275, China
* E-mail: yuhy33@mail.sysu.edu.cn

**Abstract:** We investigated the interlayer coupling effect in homobilayer structures of $M$Si$_2$X$_4$ with $M$ = Mo/W and $X$ = N/P/As. Through the combination of first-principles calculations and analytical formulations, the equilibrium interlayer distance, layer energy difference and interlayer hopping strength are obtained for all six $M$Si$_2$X$_4$ materials, which are found to be insensitive to the type of $M$ atom but differ significantly between $X$ = N and $X$ = P/As. In homobilayers with close to $0°$ twist angles, the interlayer charge redistribution introduces a stacking-dependent interlayer electrostatic potential with a magnitude reaching 0.1 eV in $M$Si$_2$N$_4$, suggesting that it can be an excellent candidate for studying the sliding ferroelectricity. The interlayer hopping strengths are found to be as large as several tens meV at valence band maxima positions $\mathbf{K}$ and $\mathbf{\Gamma}$, and ~ 1 meV at the conduction band edge $\mathbf{K}$. The resultant layer-hybridizations vary in a large range under different stacking registries, which can be used to simulate honeycomb lattice models with both trivial and non-trivial band topologies.

## I. Introduction

The discovery of graphene [1] has spurred a surge of research interest in two-dimensional (2D) layered materials, driven by their exceptional physical properties compared to their bulk counterparts as well as their enormous potential for use in miniaturized devices. Huge attentions have been paid to 2D layered semiconductors with atomically-thin thickness such as transition metal dichalcogenides (TMDs), which present a promising avenue to extend Moore's law [2]. Two monolayers of 2D materials can be vertically stacked to form a van der Waals (vdW) bilayer structure, where a moiré superlattice pattern with spatially periodic atomic stacking registries can emerge. The interlayer coupling effect, determined by the bilayer stacking registry, introduces spatially periodic modulations to potentials and layer-hybridizations experienced by carriers, which further made the moiré pattern a great platform for engineering artificial material systems and simulating novel condensed matter phases [3-6]. As a result, a series of novel quantum phenomena have been observed in bilayer TMDs systems, including various correlated electronic states [7-19], integer/fractional Chern insulators in moiré patterns [20-23] and sliding ferroelectricity in commensurate and marginally-twisted homobilayers [24-27], etc. Besides these electronic properties, monolayer and bilayer TMDs with bandgaps ranging from infrared to visible frequency have also demonstrated their potentials for next-generation nanophotonic and optoelectronic applications [3,28-31], whose optical properties are dominated by the strongly bound interlayer and intralayer excitons [32,33]. The interlayer coupling effect leads to spatially modulating electric and optical dipoles for interlayer excitons in moiré patterns [34], and splits the intralayer exciton resonance into multiple absorption peaks [35,36]. The interlayer exciton can be strongly localized by the moiré potential or an inhomogeneous electric field, which can serve as a single-photon emitter or a source of entangled photon-pairs in quantum optics [34,37]. Furthermore, by tuning the interlayer exciton energy to near the intralayer one using an out-of-plane electric field, the strong interlayer hopping can give rise to a hybridized state of the intralayer and interlayer excitons [38-43]. Such a hybrid exciton inherits the finite electric dipole from the interlayer exciton and the strong optical dipole from the intralayer one. The strong long-ranged dipole-dipole interaction between such hybrid excitons gives rise to their high nonlinearities [44], which have great potentials for applications in quantum nonlinear optics [45-47].

A natural question is whether we can find other layered semiconductors that have comparable interlayer hopping strengths with TMDs but larger interlayer separations, thus exhibit even higher nonlinearities and electrical tunability. $MoSi_2N_4$, a recently synthesized monolayer semiconductor [48], can serve as a candidate material. Monolayer $MoSi_2N_4$ and other members of the family denoted as $MSi_2X_4$ ($M$ = Mo/W, $X$ = N/P/As) have intrinsic properties comparable to or even superior to those of TMDs, including the excellent stability without a protective environment, exceptional mechanical strengths and strong absorption coefficients [48]. The corresponding crystal structure can be viewed as an inner $MX_2$ layer sandwiched between two outer $SiX$ layers, which has the same point-group symmetry ($D_{3h}$) as monolayer TMDs. As a result, the two systems are predicted to share many similar properties. For example, the lack of inversion symmetry and the presence of strong spin-orbit coupling (SOC) from transition metal $d$-orbitals result in two inequivalent valleys at the Brillouin zone corners (denoted as $\pm\mathbf{K}$). These valleys in monolayer $MSi_2X_4$ also exhibit spin-valley locking [49,50], valley-contrasting transport properties [51-53] and spin/valley dependent optical selection rules [49], similar to monolayer TMDs. The excitons in monolayer and bilayer $MSi_2X_4$ are also strongly bound with binding energies reaching several hundred meV and Bohr radii as small as 1 nm [54-57], which leads to their strong optical absorptions. This implies that $MSi_2X_4$ can serve as another candidate for applications in novel photonic and optoelectronic devices [58,59]. $MSi_2X_4$ bilayer structures offer even greater operability [60,61]. The interlayer separation of bilayer $MSi_2X_4$ (~ 1 nm) is significantly larger than bilayer TMDs (~ 0.6 nm), thus is expected to give even higher optical nonlinearities if the interlayer hopping is strong enough to hybridize intralayer and interlayer excitons. Meanwhile, first-principles calculations have pointed out a large sliding ferroelectricity can occur in AB-stacked bilayer $MoSi_2N_4$ due to a net interlayer charge transfer, implying a strong moiré potential can exist in moiré-patterned bilayer $MoSi_2N_4$ [61]. The combination of the interlayer hopping and moiré potential could lead to a spatially varying interlayer hybridization in a moiré pattern, which is essential for realizing the topologically nontrivial Chern insulator states [62,63]. Such novel quantum states have been detected in bilayer TMDs moiré patterns, while bilayer $M\text{Si}_2X_4$ could be another system to realize them. To gain a deeper understanding to these intriguing phenomena, it is imperative to conduct a thorough research on the interlayer coupling in various commensurate and moiré-patterned bilayer $M\text{Si}_2X_4$. Such works, as far as we know, have not been done yet.

Note that a moiré supercell usually contains several tens to several hundred atoms, thus it requires a very high computational cost if calculated directly from first-principles. On the other hand, the stacking registry of a local region with a size much smaller than the moiré supercell is indistinguishable from a commensurate bilayer, which can be characterized by a spatially varying interlayer translation. Commensurate bilayers thus can serve as a research platform with low computational cost for investigating the interlayer coupling of $M\text{Si}_2X_4$ bilayer moiré patterns. In this work, we employ first-principles calculations to get band structures of commensurate homobilayer $M\text{Si}_2X_4$ with various interlayer stacking registries. Combined with analytical forms that have been widely adopted for graphene and TMDs bilayers [64-68], the formulas and magnitudes of interlayer couplings in moiré patterned homobilayer $M\text{Si}_2X_4$ are obtained. Note that we focus on the homobilayer $M\text{Si}_2X_4$ system because in most heterobilayers the interlayer hopping is expected to be strongly suppressed by the large interlayer band offset.

The rest of the paper is structured as follows: the atomic models of monolayer $M\text{Si}_2X_4$ are given in Sec. II. In Sec. III, we investigated the band structures of H- and R-type homobilayer $M\text{Si}_2X_4$ based on the monolayer atomic model, and discuss the interlayer coupling induced equilibrium interlayer distances, layer energy splitting, and interlayer hopping strengths at the band extrema $\mathbf{K}$ and $\mathbf{\Gamma}$. The mini bands of moiré patterned homobilayer $M\text{Si}_2N_4$ for holes in $\mathbf{K}$ and $\mathbf{\Gamma}$ valleys and the corresponding band topologies are analyzed at the end of this section. Finally, our findings are summarized in Sec. IV.

## II. Atomic models of monolayer $M\text{Si}_2X_4$

The top and side views of the monolayer $M\text{Si}_2X_4$ atomic structure are depicted in Fig. 1(a), which forms a hexagonal lattice with a point-group of $D_{3h}$. Each monolayer can be regarded as a septuple-atomic-layer structure formed by a 2H-type $MX_2$ monolayer intercalated between two InSe-type Si$X$ monolayers. The hexagonal unit cell contains one transition metal $M$, two Si and four pnictogen atoms $X$. After optimizing the structure, the obtained lattice constants are close to 2.91 Å for $M\text{Si}_2N_4$, but become significantly larger (~ 3.5 Å) for $M\text{Si}_2P_4$ and $M\text{Si}_2As_4$, see Table I. The calculated electronic band structures of monolayer $M\text{Si}_2X_4$ based on the density functional theory (DFT) are presented in Fig. 1(c). As can be seen, monolayer $\text{MoSi}_2N_4$

(WSi$_2$N$_4$) is an indirect semiconductor with the gap value of 1.794 (2.056) eV. Their conduction band minima (CBM) are located at **K** and valence band maxima (VBM) are at **Γ**, which are hereafter denoted as **K**$_c$ and **Γ**$_v$, respectively. Meanwhile, the valence band at **K** (denoted as **K**$_v$) corresponds to a local maximum. In contrast to $M$Si$_2$N$_4$, monolayer $M$Si$_2$X$_4$ with $X$ = P/As has a direct bandgap with both VBM and CBM located at **K**. Table II indicates the orbital compositions of Bloch states at **K**$_c$, **K**$_v$ and **Γ**$_v$ for monolayer $M$Si$_2$X$_4$. The major orbital components of Bloch states at **K**$_c$ and **Γ**$_v$ are both the transition metal $d_{z^2}$ orbital, while at **K**$_v$ it is the transition metal $d_{+2} \equiv (d_{x^2-y^2} + id_{xy})/\sqrt{2}$ orbital. Compared to $M$Si$_2$N$_4$, the uplifted energy at **K**$_v$ for monolayer $M$Si$_2$P$_4$ or $M$Si$_2$As$_4$ can be attributed to the weaker hybridization between $d$-orbitals of the transition metal and $p$-orbitals of the heavier $X$ atoms [49]. Meanwhile, compared to $M$Si$_2$N$_4$, the bandgap values decrease largely for $M$Si$_2$P$_4$ and $M$Si$_2$As$_4$. Our calculated gap values summarized in Table I are in good agreement with previous works [49,50,56,69]. The alignment of CBM and VBM for different monolayers of $M$Si$_2$X$_4$ as well as TMDs is shown in Fig. 1(d), where band edge energies relative to the vacuum level are indicated. This implies that heterostructures of $M'$Si$_2$X'$_4$/$M$Si$_2$X$_4$ or $M$Si$_2$X$_4$/TMDs can have either the type-I or type-II band alignment.

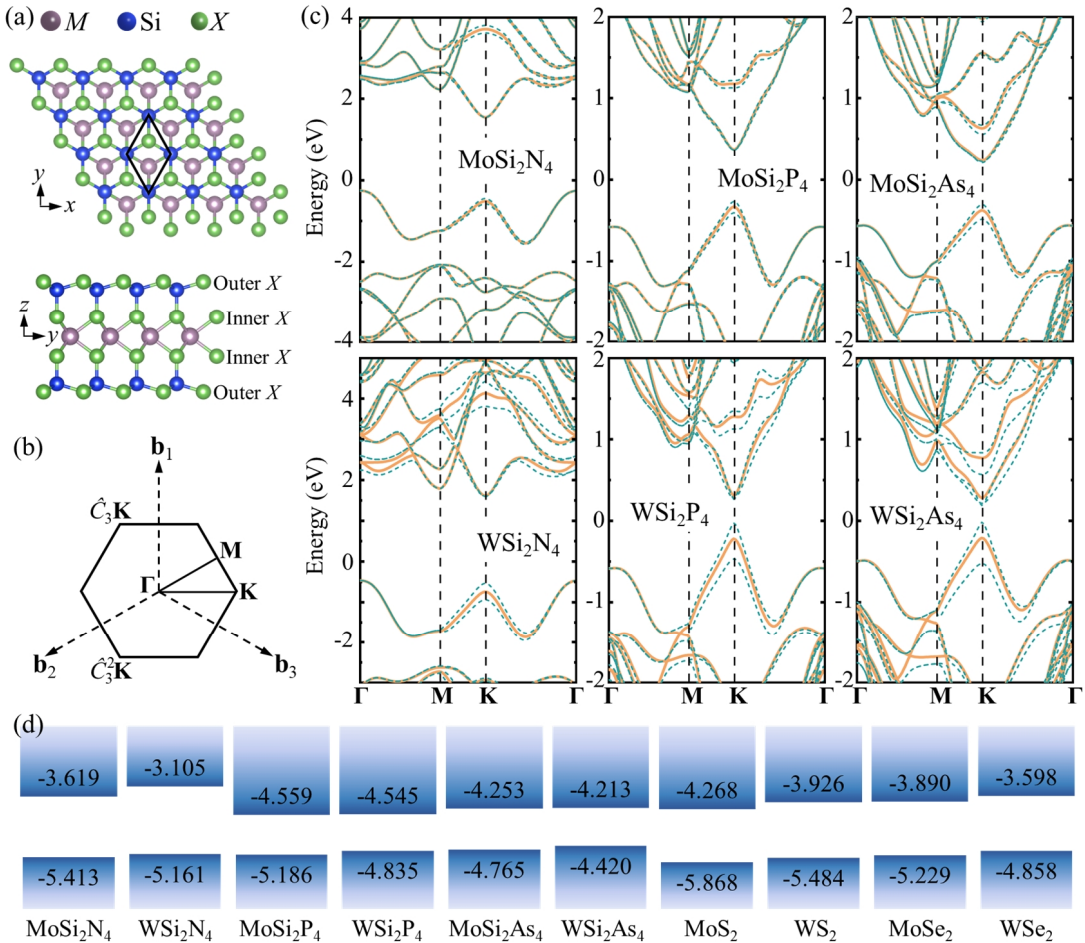

Fig. 1 (a) Top (upper) and side (lower) views of monolayer $M\text{Si}_2X_4$ crystal structure. The parallelogram indicates a primitive cell. (b) The first Brillouin zone of monolayer $M\text{Si}_2X_4$. $\Gamma$, $\mathbf{K}$ and $\mathbf{M}$ are high-symmetry points. $\mathbf{b}_{1,2,3}$ are three reciprocal lattice vectors. (c) The calculated band structures of monolayer $M\text{Si}_2X_4$ with $M$ = Mo/W and $X$ = N/P/As. The solid (dashed) curves correspond to the bands without (with) SOC. (d) The conduction/valence band edge positions for monolayer $M\text{Si}_2X_4$ and TMDs. The numbers (in the unit of eV) show conduction and valence band edge energies relative to the vacuum level.

The strong SOC from transition metal *d*-orbitals lead to a giant valence band spin splitting near $\mathbf{K}_v$, which are found to be ~ 0.13 eV for $\text{MoSi}_2X_4$ and ~ 0.40 eV for $\text{WSi}_2X_4$. This largely reduces the energy difference between $\Gamma_v$ and $\mathbf{K}_v$ in $\text{WSi}_2\text{N}_4$ compared to that without SOC. As a result, the band energy at $\Gamma_v$ is only 67.4 meV above that at $\mathbf{K}_v$ in monolayer $\text{WSi}_2\text{N}_4$, which increases to 192.1 meV in monolayer $\text{MoSi}_2\text{N}_4$. It has been suggested [49,53] that applying a strain to monolayer $M\text{Si}_2\text{N}_4$ can shift its valence band edge from $\Gamma_v$ to $\mathbf{K}_v$. The conduction band at $\mathbf{K}_c$ also has a small but finite spin splitting, whose value ranges from 3 meV in $\text{MoSi}_2\text{N}_4$ to 26 meV in $\text{WSi}_2\text{As}_4$. The calculated SOC splitting values are summarized in Table I.

Table I. The calculated lattice constants, bandgaps, SOC-induced spin splitting at $\mathbf{K}_c$ and $\mathbf{K}_v$ in monolayer $M\text{Si}_2X_4$. For comparisons, those of monolayer TMDs are also given.

|  | Lattice constant (Å) | Bandgap (eV) | $\mathbf{K}_v$ spin splitting (meV) | $\mathbf{K}_c$ spin splitting (meV) |
|---|---|---|---|---|
| $\text{MoSi}_2\text{N}_4$ | 2.910 | 1.794 | 130 | 3 |
| $\text{WSi}_2\text{N}_4$ | 2.912 | 2.056 | 404 | 11 |
| $\text{MoSi}_2\text{P}_4$ | 3.469 | 0.626 | 139 | 4 |
| $\text{WSi}_2\text{P}_4$ | 3.475 | 0.290 | 443 | 8 |
| $\text{MoSi}_2\text{As}_4$ | 3.618 | 0.512 | 181 | 16 |
| $\text{WSi}_2\text{As}_4$ | 3.623 | 0.206 | 506 | 26 |
| $\text{MoS}_2$ | 3.183 | 1.600 | 149 | 3 |
| $\text{WS}_2$ | 3.182 | 1.558 | 430 | 30 |
| $\text{MoSe}_2$ | 3.318 | 1.339 | 186 | 21 |
| $\text{WSe}_2$ | 3.316 | 1.259 | 466 | 37 |

Table II. The orbital compositions of Bloch states at **K** and **Γ** for monolayer $M\text{Si}_2X_4$. $d_{z^2}$, $d_{\pm 1} \equiv (d_{xz} \pm id_{yz})/\sqrt{2}$, $d_{\pm 2} \equiv (d_{x^2-y^2} \pm id_{xy})/\sqrt{2}$ are d-orbitals, $p_z$ and $p_{\pm 1} \equiv (p_x \pm ip_y)/\sqrt{2}$ are p-orbitals. The bold font marks the $p_z$-orbital fraction of $X$ atoms in the outer $\text{Si}_2X_2$ layers, which contributes most to the interlayer hopping strength.

| | $\mathbf{K_c}$ | $\mathbf{K_v}$ | $\mathbf{\Gamma_v}$ |
|---|---|---|---|
| MoSi$_2$N$_4$ | Mo-$d_{z^2}$ (79.4%)<br>Inner N-$p_{-1}$ (9%)<br>Mo-$s$ (3%) | Mo-$d_{+2}$ (63.2%)<br>Inner N-$p_{+1}$ (18%)<br>**Outer N-$p_z$ (5.2%)** | Mo-$d_{z^2}$ (61.9%)<br>Inner N-$p_z$ (12.4%)<br>Si-$s$ (3.4%)<br>**Outer N-$p_z$ (3.8%)**<br>Outer N-$s$ (1.8%)<br>Si-$p_z$ (1.8%) |
| WSi$_2$N$_4$ | W-$d_{z^2}$ (76.5%)<br>Inner N-$p_{-1}$ (7.6%)<br>W-$s$ (4.4%) | W-$d_{+2}$ (61.6%)<br>Inner N-$p_{+1}$ (19.4%)<br>**Outer N-$p_z$ (4.7%)** | W-$d_{z^2}$ (61.5%)<br>Inner N-$p_z$ (12.2%)<br>Si-$s$ (4%)<br>**Outer N-$p_z$ (3.6%)**<br>Outer N-$s$ (2.2%)<br>Si-$p_z$ (1.8%) |
| MoSi$_2$P$_4$ | Mo-$d_{z^2}$ (73.4%)<br>Mo-$s$ (6%)<br>Inner P-$p_{-1}$ (3.6%) | Mo-$d_{+2}$ (67%)<br>Inner P-$p_{+1}$ (11.6%)<br>**Outer P-$p_z$ (3.1%)** | Mo-$d_{z^2}$ (45.9%)<br>**Outer P-$p_z$ (8%)**<br>Si-$p_z$ (5%)<br>Inner P-$p_z$ (3%)<br>Mo-$s$ (1.9%) |
| WSi$_2$P$_4$ | W-$d_{z^2}$ (64.4%)<br>W-$s$ (8.6%)<br>Inner P-$p_{-1}$ (1.6%) | W-$d_{+2}$ (65.4%)<br>Inner P-$p_{+1}$ (12.2%)<br>**Outer P-$p_z$ (3%)** | W-$d_{z^2}$ (44.2%)<br>**Outer P-$p_z$ (8%)**<br>Si-$p_z$ (5.2%)<br>Inner P-$p_z$ (2.4%)<br>W-$s$ (2.5%) |
| MoSi$_2$As$_4$ | Mo-$d_{z^2}$ (76.9%)<br>Mo-$s$ (5%)<br>Inner As-$p_{-1}$ (3%)<br>Inner As-$d_{-1}$ (1.2%) | Mo-$d_{+2}$ (69.6%)<br>Inner As-$p_{+1}$ (11%)<br>**Outer As-$p_z$ (2.8%)** | Mo-$d_{z^2}$ (49.1%)<br>**Outer As-$p_z$ (8.2%)**<br>Si-$p_z$ (4.8%)<br>Inner As-$p_z$ (2.8%)<br>Inner As-$d_{z^2}$ (2.6%)<br>Mo-$s$ (1.5%) |
| WSi$_2$As$_4$ | W-$d_{z^2}$ (70.7%)<br>W-$s$ (7.3%)<br>Inner As-$d_{-1}$ (2%)<br>Inner As-$p_{-1}$ (1.2%) | W-$d_{+2}$ (67.8%)<br>Inner As-$p_{+1}$ (11.2%)<br>**Outer As-$p_z$ (2.8%)** | W-$d_{z^2}$ (47.1%)<br>**Outer As-$p_z$ (8.4%)**<br>Si-$p_z$ (4.8%)<br>Inner As-$d_{z^2}$ (3%)<br>Inner As-$p_z$ (1.8%)<br>W-$s$ (2%) |

## III. Interlayer couplings in R- and H-type homobilayer $MSi_2X_4$

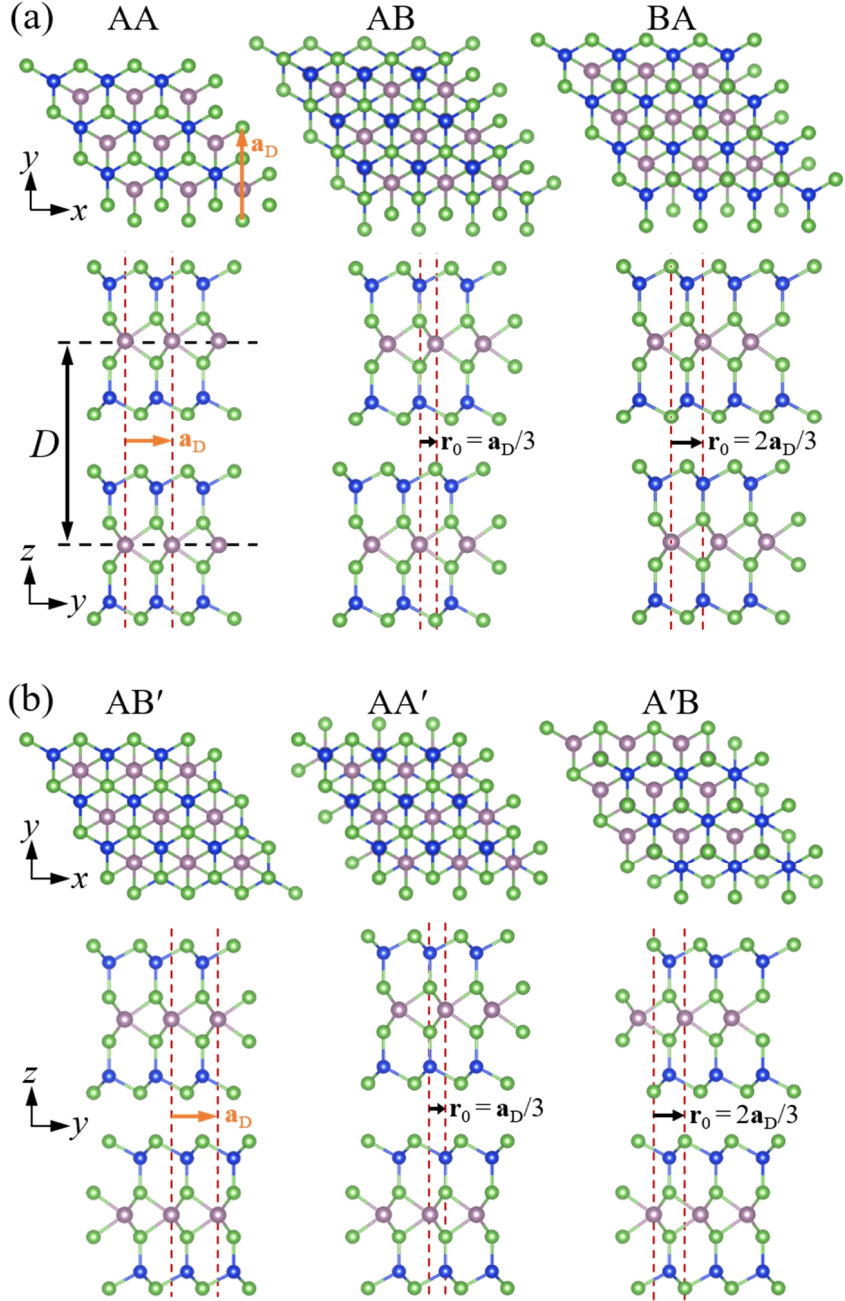

Fig. 2 (a) Top (upper) and side (lower) views of R-type high-symmetry stacking patterns AA, AB and BA for homobilayer $MSi_2X_4$, with $\mathbf{r}_0 = 0$, $\mathbf{a}_D/3$ and $2\mathbf{a}_D/3$, respectively. $\mathbf{a}_D$ corresponds to the long diagonal line of the unit cell. $D$ is the interlayer distance defined as the vertical separation between the transition metal atoms in the top and bottom layers. (b) H-type high-symmetry stacking patterns AB′, AA′ and A′B for homobilayer $MSi_2X_4$, with $\mathbf{r}_0 = 0$, $\mathbf{a}_D/3$ and $2\mathbf{a}_D/3$, respectively.

Based on the above monolayer electronic structure, we can investigate the interlayer coupling effects in bilayer structures of $MSi_2X_4$. We focus on commensurate homobilayer $MSi_2X_4$ with R- and H-type stacking patterns, which correspond to bilayers with 0° and 180° rotations between the two layers, respectively. The stacking registry of the R- or H-type bilayer can be characterized by an in-plane interlayer translation $\mathbf{r}_0$ and an interlayer separation $D$. Here $\mathbf{r}_0$ is defined as the 2D vector pointing from an $M$ atom in the lower layer to a nearby $M$ atom in the upper layer, and $D$ is the vertical distance between the two $M$-atom planes (see Fig. 2(a)). Considering the lattice periodicity, $\mathbf{r}_0$ is equivalent to $\mathbf{r}_0$ plus any lattice vector. For R-type (H-type) bilayers, we denote the high-symmetry stacking configurations with $\mathbf{r}_0 = 0$, $\mathbf{a}_D/3$ and $2\mathbf{a}_D/3$ as AA, AB and BA (AB′, AA′ and A′B), respectively, which are invariant upon an in-plane $2\pi/3$ rotation about an $M$ or Si or $X$ atom, as seen in Fig. 2. Here $\mathbf{a}_D$ corresponds to the long diagonal line of the unit cell in Fig. 1(a). The high-symmetry configurations AA, AB and BA of R-type (or AB′, AA′ and A′B of H-type) bilayers can be smoothly transformed into one another by continuously varying the translation vector $\mathbf{r}_0$. Meanwhile, an R-type bilayer can be transformed to an H-type bilayer via a 180° in-plane rotation of the upper- or lower-layer followed by a translation. For example, applying a 180° in-plane rotation centered at an $M$ atom on the upper/lower layer transforms the AA configuration to AB′. It should be emphasized that in H-type homobilayers, the upper- and lower-layers are related by an inversion. Thus the two layers in an H-type homobilayer are always equivalent. In contrast, R-type homobilayers lack inversion symmetry, and the two layers in most R-type homobilayers are not equivalent, except for AA stacking where they are related by an out-of-plane mirror reflection.

The van der Waals interaction between the layers and the resultant interlayer binding energy vary with $\mathbf{r}_0$ and $D$. For realistic commensurate bilayers, only those stacking patterns with lowest interlayer binding energies can form, which correspond to AB/BA for R-type and AB′ for H-type stackings [56]. Although R-/H-type commensurate bilayers with most $\mathbf{r}_0$ values are unstable, but local regions of moiré-patterned bilayers can be approximated by R-/H-type commensurate bilayers with spatially varying $\mathbf{r}_0$ values. For each fixed $\mathbf{r}_0$, there is an equilibrium interlayer distance $D_{eq}(\mathbf{r}_0)$ which leads to the lowest binding energy. The calculated $D_{eq}(\mathbf{r}_0)$ values for $\mathbf{r}_0$ along the long diagonal line $\mathbf{a}_D$ are shown in Fig. 3 as symbols. For the H-type homobilayer $M$Si$_2$$X_4$, the largest (smallest) $D_{eq}$ occurs at AA' with $\mathbf{r}_0 = \mathbf{a}_D/3$ (AB' with $\mathbf{r}_0 = 0$), whereas for the R-type the largest (smallest) $D_{eq}$ occurs at AA with $\mathbf{r}_0 = 0$ (AB with $\mathbf{r}_0 = \mathbf{a}_D/3$ or BA with $2\mathbf{a}_D/3$). The modulation range of $D_{eq}(\mathbf{r}_0)$ with different $\mathbf{r}_0$ is $\approx 0.5$ Å for $M$Si$_2$N$_4$, and $\approx 0.6$ Å for $M$Si$_2$P$_4$ and $M$Si$_2$As$_4$. Note that $D_{eq}(\mathbf{r}_0) \sim 10$ Å for homobilayer $M$Si$_2$$X_4$ is significantly larger than the values ($\sim 6$ Å) of homobilayer TMDs.

Meanwhile, $D_{eq}(\mathbf{r}_0)$ as a periodic function of $\mathbf{r}_0$ is found to be well fit by the below equations (shown as solid curves in Fig. 3):

$$
\begin{aligned}
D_{eq}(\mathbf{r}_0) &= D_m + \Delta D_0 |f_0(\mathbf{r}_0)|^2, &\text{(for R-type homobilayers)} \\
D_{eq}(\mathbf{r}_0) &= D_m + \Delta D_+ |f_+(\mathbf{r}_0)|^2 + \Delta D_- |f_-(\mathbf{r}_0)|^2, &\text{(for H-type homobilayers)}
\end{aligned}
\quad (1)
$$

with the periodic functions $f_{0/\pm}(\mathbf{r}_0)$ defined as

$$f_0(\mathbf{r}_0) \equiv \tfrac{1}{3}e^{i\mathbf{K}\cdot\mathbf{r}_0}\left(e^{-i\mathbf{K}\cdot\mathbf{r}_0} + e^{-i\hat{C}_3\mathbf{K}\cdot\mathbf{r}_0} + e^{-i\hat{C}_3^2\mathbf{K}\cdot\mathbf{r}_0}\right),$$
$$f_\pm(\mathbf{r}_0) \equiv \tfrac{1}{3}e^{i\mathbf{K}\cdot\mathbf{r}_0}\left(e^{-i\mathbf{K}\cdot\mathbf{r}_0} + e^{-i(\hat{C}_3\mathbf{K}\cdot\mathbf{r}_0 \pm 2\pi/3)} + e^{-i(\hat{C}_3^2\mathbf{K}\cdot\mathbf{r}_0 \pm 4\pi/3)}\right). \quad (2)$$

In Eq. (1), $D_\mathrm{m} = \min\left(D_\mathrm{eq}(\mathbf{r}_0)\right)$ is the minimum equilibrium interlayer distance for all $\mathbf{r}_0$ values. $f_{0/\pm}(\mathbf{r}_0)$ satisfy $|f_0(\mathbf{r}_0)|^2 + |f_+(\mathbf{r}_0)|^2 + |f_-(\mathbf{r}_0)|^2 = 1$, $f_0(0) = f_+(\mathbf{a}_D/3) = f_-(2\mathbf{a}_D/3) = 1$ and $f_0(\mathbf{a}_D/3) = f_0(2\mathbf{a}_D/3) = f_\pm(0) = f_+(2\mathbf{a}_D/3) = f_-(\mathbf{a}_D/3) = 0$. The fitting parameters $\Delta D_{0/\pm}$ for $M\mathrm{Si}_2X_4$ homobilayers are summarized in Table III.

Table III. Fitting parameters for the curves in Fig. 3 using Eq. (1).

|        |                  | MoSi$_2$N$_4$ | WSi$_2$N$_4$ | MoSi$_2$P$_4$ | WSi$_2$P$_4$ | MoSi$_2$As$_4$ | WSi$_2$As$_4$ |
|--------|------------------|---------------|--------------|---------------|--------------|----------------|---------------|
| R-type | $D_\mathrm{m}$ (Å) | 10.065 | 10.058 | 12.577 | 12.571 | 13.196 | 13.177 |
|        | $\Delta D_0$ (Å)   | 0.456  | 0.454  | 0.666  | 0.665  | 0.618  | 0.644  |
| H-type | $D_\mathrm{m}$ (Å) | 9.976  | 9.967  | 12.558 | 12.554 | 13.167 | 13.155 |
|        | $\Delta D_+$ (Å)   | 0.514  | 0.518  | 0.668  | 0.682  | 0.664  | 0.661  |
|        | $\Delta D_-$ (Å)   | 0.243  | 0.234  | 0.120  | 0.120  | 0.095  | 0.103  |

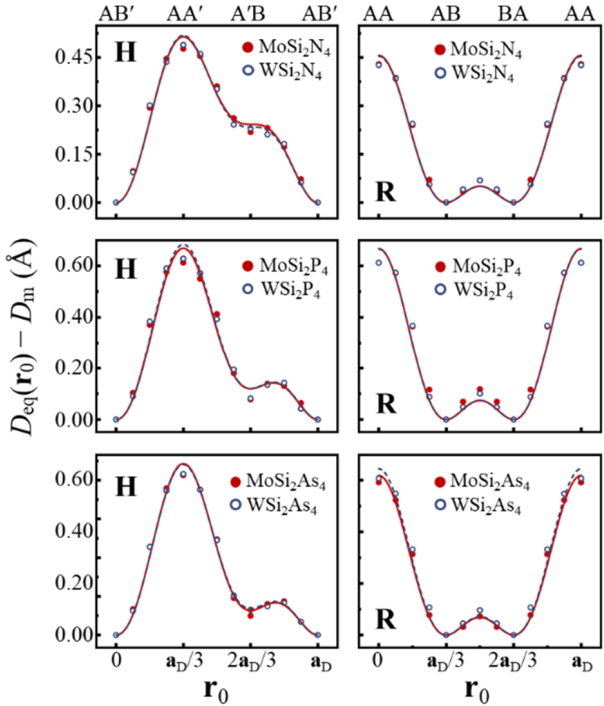

Fig. 3 The equilibrium interlayer separations $D_\mathrm{eq}(\mathbf{r}_0)$ of H-type (left panels) and R-type (right panels) homobilayer $M\mathrm{Si}_2X_4$ as functions of $\mathbf{r}_0$. Solid curves are fittings using Eq. (1).

Fig. 4(a-c) show our calculated band structures of homobilayer $M\mathrm{Si}_2X_4$ under high-symmetry stacking configurations in the absence of SOC. For each stacking registry characterized by $\mathbf{r}_0$, the interlayer distance $D$ is set as the corresponding equilibrium value $D_\mathrm{eq}(\mathbf{r}_0)$. Compared to the layer-decoupled case with doubly degenerate conduction/valence bands from the layer degree-of-freedom (spin not considered), the finite interlayer coupling

effect gives rise to two layer-hybridized conduction/valence sub-bands with different energies. Here we focus on sub-bands near the band extrema $\mathbf{K}_c$, $\mathbf{K}_v$ and $\mathbf{\Gamma}_v$ whose energies are well separated from other bands. At these momentum-space positions, the relative alignment of the two sub-bands differs between stacking configurations, indicating that the interlayer coupling effect varies with $\mathbf{r}_0$.

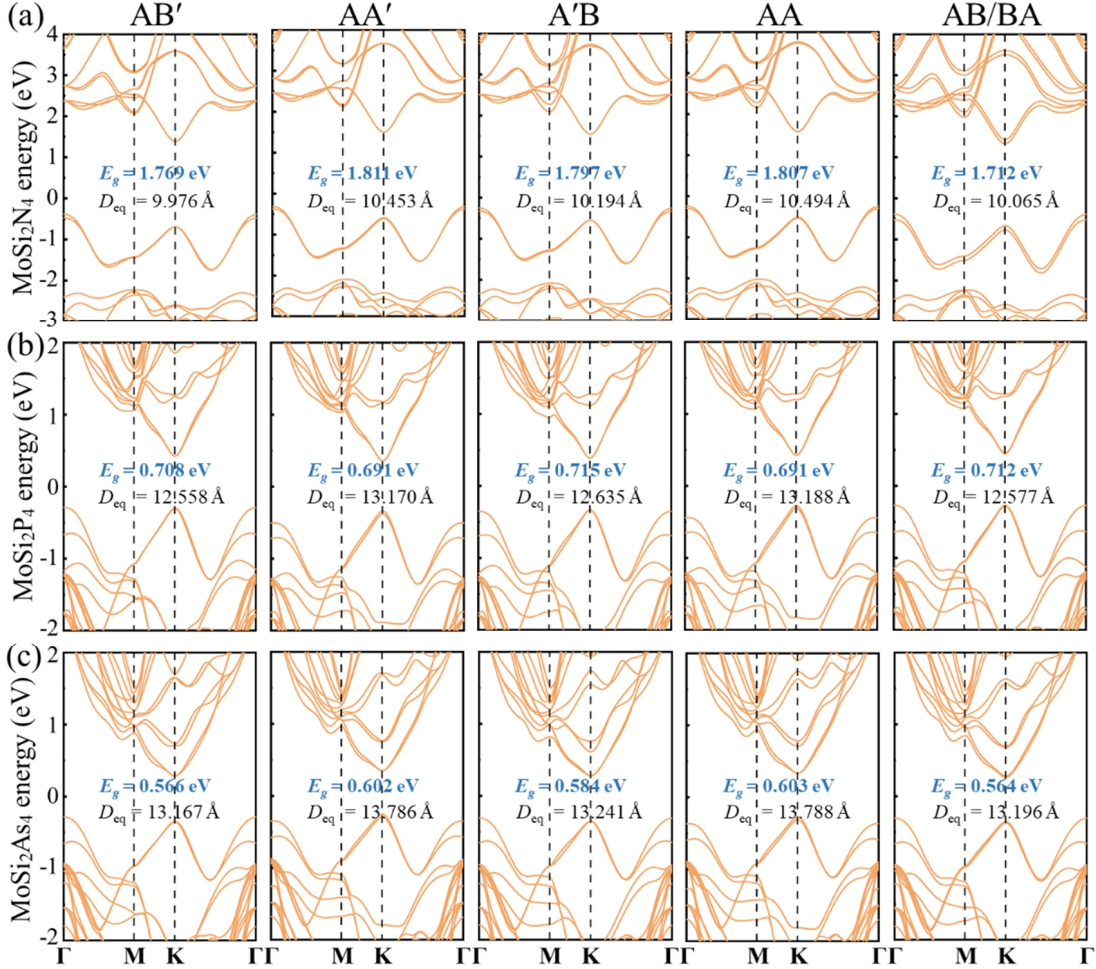

Fig. 4 The calculated band structures of homobilayer (a) $MoSi_2N_4$, (b) $MoSi_2P_4$ and (c) $MoSi_2As_4$ with high-symmetry stackings configurations AB', AA', A'B, AA and AB/BA, under the corresponding equilibrium interlayer distance $D_{eq}$.

For theoretical insights, below we set $\mathbf{r}_0$ and $D$ as independent parameters which can vary freely. To quantify the interlayer coupling induced layer-hybridization, we start from the monolayer Bloch states $|n, \mathbf{k}\rangle$ and $|n', \mathbf{k}\rangle$ with $n$ = c/v ($n'$ = c'/v') referring to the conduction/valence band of the lower (upper) layer. In the subspace spanning the two closely aligned sub-bands, the R/H-type bilayer Hamiltonian can be well described by a 2×2 matrix

$$\hat{H}_{n,\mathbf{k}}^{(R/H)} = E_{n,\mathbf{k}}^{(R/H)} + \delta E_{n,\mathbf{k}}^{(R/H)} \hat{\sigma}_z + T_{n,\mathbf{k}}^{(R/H)} \hat{\sigma}_+ + T_{n,\mathbf{k}}^{(R/H)*} \hat{\sigma}_-. \tag{3}$$

Here $\hat{\sigma}_z \equiv |n, \mathbf{k}\rangle\langle n, \mathbf{k}| - |n', \mathbf{k}\rangle\langle n', \mathbf{k}|$, $\hat{\sigma}_+ \equiv |n, \mathbf{k}\rangle\langle n', \mathbf{k}|$ and $\hat{\sigma}_- \equiv |n', \mathbf{k}\rangle\langle n, \mathbf{k}|$ are the Pauli matrices for the layer-pseudospin. $E_{n,\mathbf{k}}^{(R/H)} \pm \delta E_{n,\mathbf{k}}^{(R/H)}$ account for the diagonal energies of basis states $|n,\mathbf{k}\rangle$ and $|n',\mathbf{k}\rangle$, and $T_{n,\mathbf{k}}^{(R/H)}$ corresponds to their interlayer hopping. Generally, $E_{n,\mathbf{k}}^{(R/H)} = E_{n,\mathbf{k}}^{(R/H)}(D,\mathbf{r}_0)$, $\delta E_{n,\mathbf{k}}^{(R/H)} = \delta E_{n,\mathbf{k}}^{(R/H)}(D,\mathbf{r}_0)$ and $T_{n,\mathbf{k}}^{(R/H)} = T_{n,\mathbf{k}}^{(R/H)}(D,\mathbf{r}_0)$ all decay with $D$ and vary periodically with $\mathbf{r}_0$. The eigen-states of $\hat{H}_{n,\mathbf{k}}^{(R/H)}$ correspond to two layer-hybridized sub-bands, with eigen-energies $\mathcal{E}_{n\pm,\mathbf{k}}^{(R/H)}(D,\mathbf{r}_0)$ given by

$$\mathcal{E}_{n\pm,\mathbf{k}}^{(R/H)}(D,\mathbf{r}_0) = E_{n,\mathbf{k}}^{(R/H)}(D,\mathbf{r}_0) \pm \sqrt{\left|\delta E_{n,\mathbf{k}}^{(R/H)}(D,\mathbf{r}_0)\right|^2 + \left|T_{n,\mathbf{k}}^{(R/H)}(D,\mathbf{r}_0)\right|^2}. \quad (4)$$

The energy splitting $\Delta \mathcal{E}_{n,\mathbf{k}}^{(R/H)}(D,\mathbf{r}_0) \equiv \mathcal{E}_{n+,\mathbf{k}}^{(R/H)}(D,\mathbf{r}_0) - \mathcal{E}_{n-,\mathbf{k}}^{(R/H)}(D,\mathbf{r}_0)$ between the two sub-bands is

$$\Delta \mathcal{E}_{n,\mathbf{k}}^{(R/H)}(D,\mathbf{r}_0) = 2\sqrt{\left|\delta E_{n,\mathbf{k}}^{(R/H)}(D,\mathbf{r}_0)\right|^2 + \left|T_{n,\mathbf{k}}^{(R/H)}(D,\mathbf{r}_0)\right|^2}. \quad (5)$$

Below we focus on $\mathbf{k}$ located exactly at the conduction and valence band extrema $\mathbf{K}_c$, $\mathbf{K}_v$ and $\mathbf{\Gamma}_v$, where symmetries enforce $E_{n,\mathbf{k}}^{(R/H)}(D,\mathbf{r}_0)$, $\delta E_{n,\mathbf{k}}^{(R/H)}(D,\mathbf{r}_0)$ and $T_{n,\mathbf{k}}^{(R/H)}(D,\mathbf{r}_0)$ to have specific forms [68].

Note that any periodic function $f(\mathbf{r}_0)$ can be Fourier expanded as $f(\mathbf{r}_0) = \sum_{\mathbf{G}} F(\mathbf{G}) e^{i\mathbf{G}\cdot\mathbf{r}_0}$. For real functions $\delta E_{n,\mathbf{k}}^{(R/H)}(D,\mathbf{r}_0)$ and $E_{n,\mathbf{k}}^{(R/H)}(D,\mathbf{r}_0)$ which vary smoothly with $\mathbf{r}_0$, only the main Fourier components with $\mathbf{G} = 0$, $\pm\mathbf{b}_1$, $\pm\mathbf{b}_2$ and $\pm\mathbf{b}_3$ need to be kept ($\pm\mathbf{b}_{1,2,3}$ are nonzero reciprocal lattice vectors with the smallest magnitude, see Fig. 1(b)). Meanwhile the following symmetry properties should be considered: (1) $\delta E_{n,\mathbf{K}/\mathbf{\Gamma}}^{(R/H)}(D,\hat{C}_3\mathbf{r}_0) = \delta E_{n,\mathbf{K}/\mathbf{\Gamma}}^{(R/H)}(D,\mathbf{r}_0)$ and $E_{n,\mathbf{K}/\mathbf{\Gamma}}^{(R/H)}(D,\hat{C}_3\mathbf{r}_0) = E_{n,\mathbf{K}/\mathbf{\Gamma}}^{(R/H)}(D,\mathbf{r}_0)$ due to the $\hat{C}_3$-symmetry at $\mathbf{K}$ and $\mathbf{\Gamma}$. (2) In R-type homobilayers, applying an out-of-plane mirror reflection switches the two layers and changes the interlayer translation from $\mathbf{r}_0$ to $-\mathbf{r}_0$, which results in $\delta E_{n,\mathbf{k}}^{(R)}(D,-\mathbf{r}_0) = -\delta E_{n,\mathbf{k}}^{(R)}(D,\mathbf{r}_0)$ and $E_{n,\mathbf{k}}^{(R)}(D,-\mathbf{r}_0) = E_{n,\mathbf{k}}^{(R)}(D,\mathbf{r}_0)$ for any $\mathbf{k}$ value. (3) In H-type homobilayers without including SOC, the two layers are related by an inversion so $\delta E_{n,\mathbf{k}}^{(H)}(D,\mathbf{r}_0) = 0$. These constrains enforce $E_{n,\mathbf{K}/\mathbf{\Gamma}}^{(R/H)}(D,\mathbf{r}_0)$ and $\delta E_{n,\mathbf{K}/\mathbf{\Gamma}}^{(R/H)}(D,\mathbf{r}_0)$ to be in the following forms

$$\begin{aligned}
E_{n,\mathbf{K}/\mathbf{\Gamma}}^{(R)}(D,\mathbf{r}_0) &= \delta_{n,\mathbf{K}/\mathbf{\Gamma}}^{(R)}(D)[|f_+(\mathbf{r}_0)|^2 + |f_-(\mathbf{r}_0)|^2] + \text{Const.}, \\
\delta E_{n,\mathbf{K}/\mathbf{\Gamma}}^{(R)}(D,\mathbf{r}_0) &= \Delta_{n,\mathbf{K}/\mathbf{\Gamma}}^{(R)}(D)[|f_+(\mathbf{r}_0)|^2 - |f_-(\mathbf{r}_0)|^2], \\
E_{n,\mathbf{K}/\mathbf{\Gamma}}^{(H)}(D,\mathbf{r}_0) &= \delta_{n,\mathbf{K}/\mathbf{\Gamma}}^{(H)}(D)|f_+(\mathbf{r}_0)|^2 + \Delta_{n,\mathbf{K}/\mathbf{\Gamma}}^{(H)}(D)|f_-(\mathbf{r}_0)|^2 + \text{Const.}, \\
\delta E_{n,\mathbf{K}/\mathbf{\Gamma}}^{(H)}(D,\mathbf{r}_0) &= 0.
\end{aligned} \quad (6)$$

Meanwhile, the general form of the complex interlayer hopping $T_{n,\mathbf{k}}^{(R/H)}(D,\mathbf{r}_0) = \sum_{\mathbf{G}} \tilde{T}_n^{(R/H)}(D,\mathbf{k}+\mathbf{G}) e^{-i\mathbf{G}\cdot\mathbf{r}_0}$ has been obtained in early works [64-68]. Here $\tilde{T}_n^{(R/H)}(D,\mathbf{k}+\mathbf{G})$ is the Fourier transform of the hopping integral between two localized orbitals, whose absolute value decreases quickly with the increase of $|\mathbf{k} + \mathbf{G}|$ thus only a few $\tilde{T}_n^{(R/H)}(D, \mathbf{k} + \mathbf{G})$ terms need to be kept. At $\mathbf{K}_c$ and $\mathbf{K}_v$, in most cases we only need to keep the main terms with $\mathbf{K} + \mathbf{G} = \mathbf{K}, \hat{C}_3\mathbf{K}, \hat{C}_3^2\mathbf{K}$; for the valence band in H-type homobilayers, higher-order terms with $\mathbf{K} + \mathbf{G} = -2\mathbf{K}, -2\hat{C}_3\mathbf{K}, -2\hat{C}_3^2\mathbf{K}$ and $\kappa_1 \equiv \mathbf{K} - \mathbf{b}_2, \kappa_2 \equiv \mathbf{K} + \mathbf{b}_3, \hat{C}_3\kappa_1, \hat{C}_3\kappa_2, \hat{C}_3^2\kappa_1, \hat{C}_3^2\kappa_2$ can also be included to give better fitting accuracies. On the other hand, at $\Gamma_v$ we only need to keep the $\mathbf{G} = 0$ term. Considering the $\hat{C}_3$ symmetry and orbital compositions at $\mathbf{K}$, the interlayer hopping form can be simplified to

$$T_{n,\mathbf{K}}^{(R)}(D, \mathbf{r}_0) = t_{n,\mathbf{K}}^{(R)}(D) f_0(\mathbf{r}_0),$$

$$T_{c,\mathbf{K}}^{(H)}(D, \mathbf{r}_0) = t_{c,\mathbf{K}}^{(H)}(D) f_0(\mathbf{r}_0),$$

$$T_{v,\mathbf{K}}^{(H)}(D, \mathbf{r}_0) = t_{v,\mathbf{K}}^{(H)}(D) f_+(\mathbf{r}_0) + t_{v,\mathbf{K}}'^{(H)}(D) g_+(\mathbf{r}_0) \qquad (7)$$

$$+ t_{v,\mathbf{K}}''^{(H)}(D) g_+^{(1)}(\mathbf{r}_0) + t_{v,\mathbf{K}}''^{(H)*}(D) g_+^{(2)}(\mathbf{r}_0),$$

$$T_{v,\Gamma}^{(R/H)}(D, \mathbf{r}_0) = t_{v,\Gamma}^{(R/H)}(D),$$

with $f_0(\mathbf{r}_0)$ and $f_+(\mathbf{r}_0)$ given in Eq. (2) and $g_+(\mathbf{r}_0) \equiv \frac{1}{3}e^{i\mathbf{K}\cdot\mathbf{r}_0}\left(e^{2i\mathbf{K}\cdot\mathbf{r}_0} + e^{i(2\hat{C}_3\mathbf{K}\cdot\mathbf{r}_0 - 2\pi/3)} + e^{i(2\hat{C}_3^2\mathbf{K}\cdot\mathbf{r}_0 - 4\pi/3)}\right)$, $g_+^{(1,2)}(\mathbf{r}_0) \equiv \frac{1}{3}e^{i\mathbf{K}\cdot\mathbf{r}_0}\left(e^{i\kappa_{1,2}\cdot\mathbf{r}_0} + e^{i(\hat{C}_3\kappa_{1,2}\cdot\mathbf{r}_0 - 2\pi/3)} + e^{i(\hat{C}_3^2\kappa_{1,2}\cdot\mathbf{r}_0 - 4\pi/3)}\right)$. Note that $t_{n,\mathbf{K}}^{(R/H)}(D) \equiv \tilde{T}_n^{(R/H)}(D, \mathbf{K})$ corresponds to the main hopping term, whereas $t_{v,\mathbf{K}}'^{(H)}(D) \equiv \tilde{T}_v^{(H)}(D, -2\mathbf{K})$ and $t_{v,\mathbf{K}}''^{(H)}(D) \equiv \tilde{T}_v^{(H)}(D, \kappa_1)$ are higher-order hopping terms with $|t_{n,\mathbf{K}}^{(H)}(D)| \gg |t_{n,\mathbf{K}}'^{(H)}(D)|, |t_{n,\mathbf{K}}''^{(H)}(D)|$.

**(a) Interlayer couplings at $\mathbf{K}_c$ and $\mathbf{K}_v$**

From Eq. (4-7), we can write the energies of the two sub-bands at $\mathbf{K}$ as

$$\mathcal{E}_{n\pm,\mathbf{K}}^{(R)}(D, \mathbf{r}_0) = E_{n,\mathbf{K}}^{(R)} + \delta_{n,\mathbf{K}}^{(R)}(D)[|f_+(\mathbf{r}_0)|^2 + |f_-(\mathbf{r}_0)|^2]$$

$$\pm \sqrt{\left(\Delta_{n,\mathbf{K}}^{(R)}(D)[|f_+(\mathbf{r}_0)|^2 - |f_-(\mathbf{r}_0)|^2]\right)^2 + \left|t_{n,\mathbf{K}}^{(R)}(D) f_0(\mathbf{r}_0)\right|^2},$$

$$\mathcal{E}_{n\pm,\mathbf{K}}^{(H)}(D, \mathbf{r}_0) = E_{n,\mathbf{K}}^{(H)} + \delta_{n,\mathbf{K}}^{(H)}(D)|f_+(\mathbf{r}_0)|^2 + \Delta_{n,\mathbf{K}}^{(H)}(D)|f_-(\mathbf{r}_0)|^2 \qquad (8)$$

$$\pm \Big| t_{v,\mathbf{K}}^{(H)}(D) f_+(\mathbf{r}_0) + t_{v,\mathbf{K}}'^{(H)}(D) g_+(\mathbf{r}_0) + t_{v,\mathbf{K}}''^{(H)}(D) g_+^{(1)}(\mathbf{r}_0)$$

$$+ t_{v,\mathbf{K}}''^{(H)*}(D) g_+^{(2)}(\mathbf{r}_0) \Big|.$$

We first set the interlayer separation $D$ to its minimum value $D_m$ given in Table III for the corresponding R-/H-type homobilayer material. By fitting the calculated $\mathcal{E}_{n\pm,\mathbf{K}}^{(R/H)}(D_m, \mathbf{r}_0)$ under various $\mathbf{r}_0$ values to Eq. (8), we can get the parameters $\delta_{n,\mathbf{K}}^{(R/H)}(D_m)$, $\Delta_{n,\mathbf{K}}^{(R/H)}(D_m)$ and $t_{n,\mathbf{K}}^{(R/H)}(D_m)$ which quantify the interlayer coupling effect.

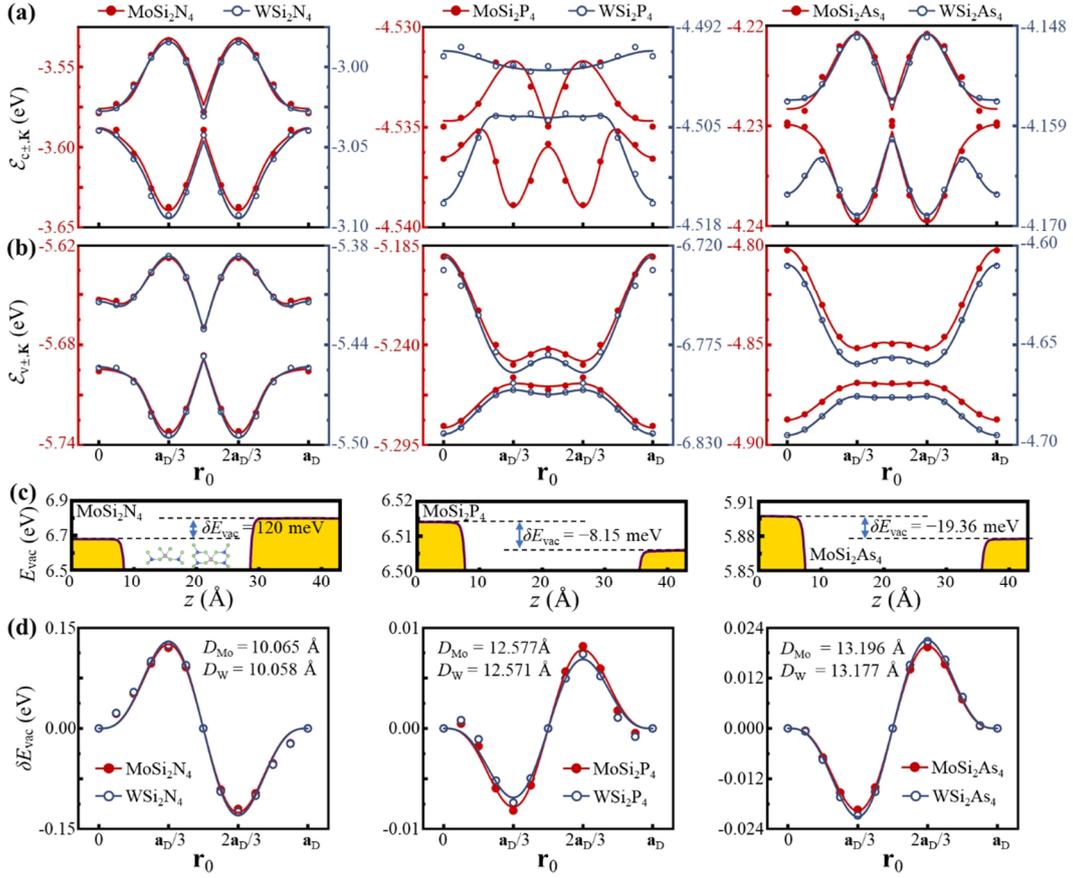

Fig. 5. (a) The conduction band energies $\mathcal{E}^{(R)}_{c\pm,K}(D_m, r_0)$ at **K** with respect to the vacuum level as functions of $r_0$ for R-type homobilayer (left) $M$Si$_2$N$_4$, (middle) $M$Si$_2$P$_4$ and (right) $M$Si$_2$As$_4$. The symbols correspond to the numerically calculated band energies, and the curves are analytical fittings using Eq. (4) with the fitting parameters given in Table IV. (b) The valence band energies $\mathcal{E}^{(R)}_{v\pm,K}(D_m, r_0)$ at **K**. (c) The vacuum energies $E_{vac}$ in AB-stacked ($r_0 = a_D/3$) homobilayers as a function of $z$ coordinate. (d) The vacuum energy differences $\delta E_{vac}$ between the two sides of homobilayers as functions of $r_0$. In (a-d), the corresponding minimum interlayer distances $D_m$ are $D_{Mo} = 10.065$ Å, 12.577 Å and 13.196 Å for MoSi$_2$N$_4$, MoSi$_2$P$_4$ and MoSi$_2$As$_4$, respectively, and $D_W = 10.058$ Å, 12.571 Å and 13.177 Å for WSi$_2$N$_4$, WSi$_2$P$_4$ and WSi$_2$As$_4$, respectively.

In Fig. 5, the symbols correspond to the numerically calculated eigen-energies $\mathcal{E}^{(R)}_{c\pm,K}$ and $\mathcal{E}^{(R)}_{v\pm,K}$ for R-type homobilayers as functions of $r_0$ under $D_m$, and the curves are analytical fittings using Eq. (8). We can see that the analytical forms agree quite well with numerical results. The fitting results are summarized in Table IV, which indicate that in R-type homobilayer $M$Si$_2$X$_4$ the modulation of $\mathcal{E}^{(R)}_{n\pm,K}(D_m, r_0)$ with $r_0$ is dominated by either $t^{(R)}_{n,K}$ or $\Delta^{(R)}_{n,K}$, whereas the effect of $\delta^{(R)}_{n,K}$ is weak. Interestingly, values of $t^{(R)}_{v,K}$ are nearly the same for MoSi$_2$X$_4$ and WSi$_2$X$_4$ with the same $X$ atom, but differ by an order of magnitude between $X$ = N and $X$ = P/As. In fact, although the main composition of the Bloch state at $K_{c/v}$ is $d_{z^2}/d_{+2}$-orbital of $M$ atoms, the corresponding interlayer hopping $t^{(R)}_{c/v,K}$ is mainly determined by the small fraction of $X$ atoms in the outer Si$_2$X$_2$ layers (denoted as outer $X$). From the orbital compositions in Table II, we can see that the strength of $t_{n,\mathbf{K}}^{(R)}$ is directly related to the $p_z$-orbital fraction of outer $X$ atoms. For instance, in MoSi$_2$X$_4$ and WSi$_2$X$_4$ with the same $X$ atom, their Bloch states at $\mathbf{K}_v$ have nearly the same $p_z$-orbital fractions of outer $X$ atoms (~ 5% in $M$Si$_2$N$_4$, ~ 3% in $M$Si$_2$P$_4$ and $M$Si$_2$As$_4$), resulting in rather close values of $t_{v,\mathbf{K}}^{(R)}$. Meanwhile, the $p_z$-orbital fraction of outer $X$ atoms vanishes at $\mathbf{K}_c$ due to the $\hat{C}_3$-symmetry constrain, thus the strength of $t_{c,\mathbf{K}}^{(R)}$ is much weaker.

Table IV. Fitting parameters for the curves in Fig. 5 and Fig. 6 using Eq. (4-7). The units are meV.

|  | MoSi$_2$N$_4$ | WSi$_2$N$_4$ | MoSi$_2$P$_4$ | WSi$_2$P$_4$ | MoSi$_2$As$_4$ | WSi$_2$As$_4$ |
|---|---|---|---|---|---|---|
| $\delta_{c,\mathbf{K}}^{(R)}(D_m)$ | -1.990 | -4.390 | 0.448 | 4.380 | -0.393 | 2.350 |
| $\|\Delta_{c,\mathbf{K}}^{(R)}(D_m)\|$ | 53.461 | 55.360 | 3.328 | 2.960 | 9.531 | 9.660 |
| $t_{c,\mathbf{K}}^{(R)}(D_m)$ | 7.018 | 7.726 | 0.720 | 8.940 | 0.520 | 4.787 |
| $\delta_{v,\mathbf{K}}^{(R)}(D_m)$ | -5.470 | -5.930 | -16.430 | -16.550 | -15.500 | -14.910 |
| $\|\Delta_{v,\mathbf{K}}^{(R)}(D_m)\|$ | 53.221 | 55.150 | 2.873 | 1.520 | 7.550 | 6.880 |
| $t_{v,\mathbf{K}}^{(R)}(D_m)$ | 22.170 | 20.620 | 43.574 | 42.120 | 39.714 | 39.590 |
| $\Delta_{vac}(D_m)$ | 125.180 | 129.530 | -7.780 | -6.850 | -19.610 | -21.010 |
| $t_{v,\mathbf{K}}^{(H)}(D_m)$ | 22.996 | 20.888 | 44.971 | 40.668 | 38.495 | 37.963 |
| $t'_{v,\mathbf{K}}^{(H)}(D_m)$ | 0 | 0 | 5.365 | 3.225 | 3.390 | 3.416 |
| $\|t''_{v,\mathbf{K}}^{(H)}(D_m)\|$ | 0 | 0 | 5.255 | 1.111 | 1.220 | 1.184 |

$\delta E_{n,\mathbf{K}}^{(R)}(D,\mathbf{r}_0)$ accounts for the diagonal energy difference between the two basis Bloch states in two different monolayers (see Eq. (3)). It has been pointed out that in bilayer TMDs, such an interlayer energy difference is induced predominantly by the spontaneous electrical polarization across the bilayer interface, i.e., an interlayer charge redistribution effect[70]. This property also applies to homobilayer $M$Si$_2$X$_4$. Fig. 5(d) shows vacuum energies $E_{vac}$ in AB-stacked ($\mathbf{r}_0 = \mathbf{a}_D/3$) homobilayers as a function of $z$ coordinate, where the difference $\delta E_{vac}$ at the top and bottom sides of the homobilayer comes from a built-in voltage induced by the interlayer charge redistribution. Generally $\delta E_{vac}$ varies with both $D$ and $\mathbf{r}_0$, which from our calculation can be well fit by $\delta E_{vac}(D,\mathbf{r}_0) = \Delta_{vac}(D)[|f_+(\mathbf{r}_0)|^2 - |f_-(\mathbf{r}_0)|^2]$ (see Fig. 5(d)). The fitting result indicates $\Delta_{vac}(D) \approx 2\Delta_{c,\mathbf{K}}^{(R)}(D) \approx 2\Delta_{v,\mathbf{K}}^{(R)}(D)$, in agreement with the prediction that $\delta E_{vac}(D,\mathbf{r}_0) \approx 2\delta E_{n,\mathbf{K}}^{(R)}(D,\mathbf{r}_0)$. Note that $\delta E_{vac}$ of $M$Si$_2$N$_4$ has a much larger magnitude and an opposite sign compared to those of $M$Si$_2$P$_4$ and $M$Si$_2$As$_4$. This suggests that $M$Si$_2$N$_4$ can be an excellent candidate for studying the recently emerged sliding ferroelectricity, which has been experimentally reported in hexagonal boron nitride and TMDs [24-27,71-74].

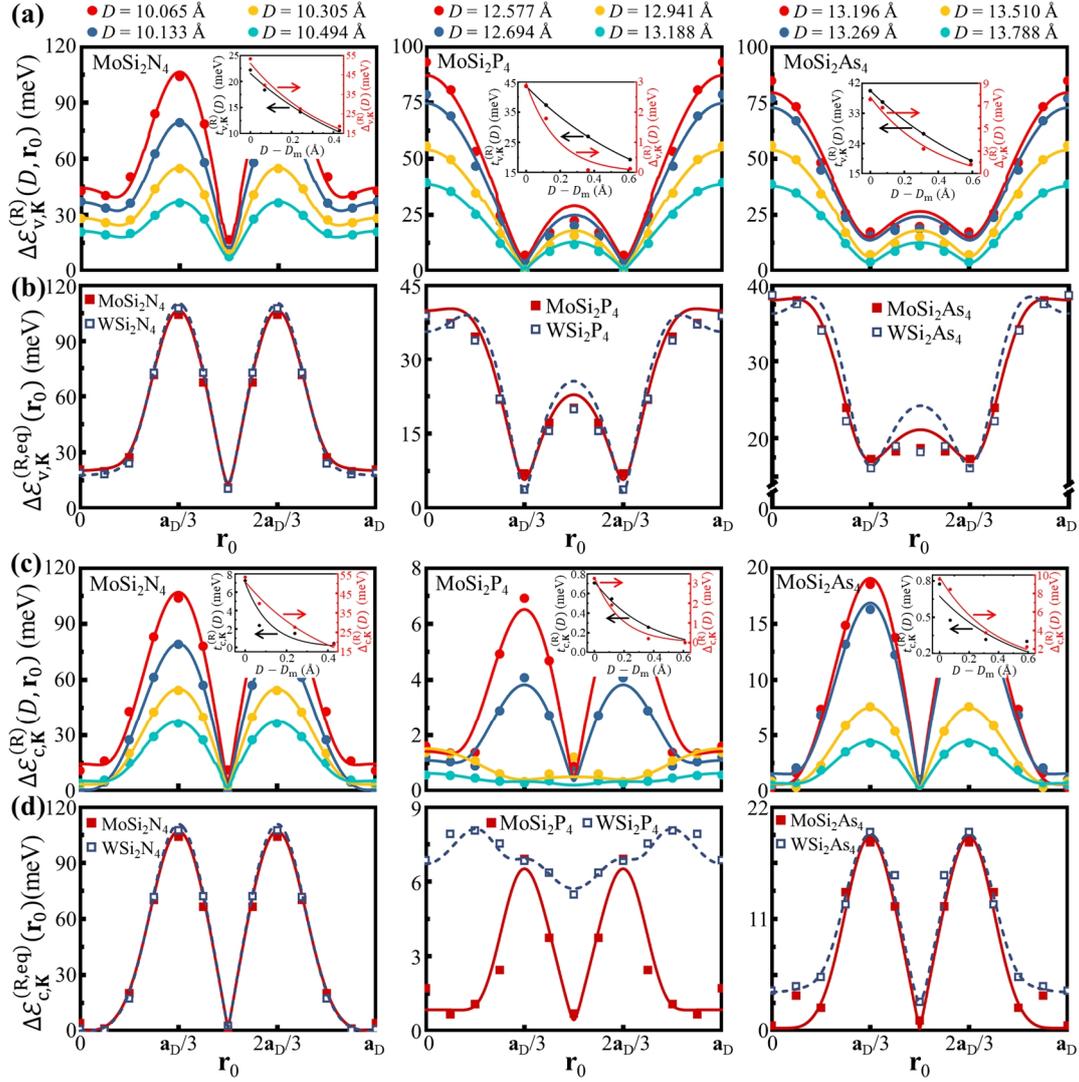

Fig. 6 (a) The energy splitting $\Delta\mathcal{E}_{v,\mathbf{K}}^{(R)}(D,\mathbf{r}_0)$ between the two valence sub-bands at $\mathbf{K}$ in R-type homobilayer MoSi$_2$X$_4$, under four different values of the interlayer distance $D$. (b) The valence sub-band splitting $\Delta\mathcal{E}_{v,\mathbf{K}}^{(R,eq)}(\mathbf{r}_0)$ under the equilibrium interlayer distance $D_{eq}(\mathbf{r}_0)$ for MoSi$_2$X$_4$ and WSi$_2$X$_4$. (c) The energy splitting $\Delta\mathcal{E}_{c,\mathbf{K}}^{(R)}(D,\mathbf{r}_0)$ between the two conduction sub-bands at $\mathbf{K}$ in R-type homobilayer MoSi$_2$X$_4$. (d) The conduction sub-band splitting $\Delta\mathcal{E}_{c,\mathbf{K}}^{(R,eq)}(\mathbf{r}_0)$ under the equilibrium interlayer distance. In (a-d), symbols correspond to the numerically calculated values, and curves are analytical fittings using Eq. (8) and Eq. (10).

The interlayer coupling depends sensitively on the interlayer distance $D$. The strengths of $t_{n,\mathbf{K}}^{(R/H)}$ and $\Delta_{n,\mathbf{K}}^{(R/H)}$ are expected to decay exponentially with the increase of $D$. Fig. 6(a,c) show values of $\Delta\mathcal{E}_{n,\mathbf{K}}^{(R)}(D,\mathbf{r}_0) = 2\sqrt{\left|\delta E_{n,\mathbf{K}}^{(R)}(D,\mathbf{r}_0)\right|^2 + \left|T_{n,\mathbf{K}}^{(R)}(D,\mathbf{r}_0)\right|^2}$ as functions of $\mathbf{r}_0$ in R-type homobilayer $M$Si$_2$X$_4$ under four different $D$ values. Numerical values of $\Delta\mathcal{E}_{n,\mathbf{K}}^{(R)}(D,\mathbf{r}_0)$ are shown as solid symbols, which can be fit rather well with Eq. (8) (shown as solid curves). The fitting parameters $t_{n,\mathbf{K}}^{(R)}(D)$ and $\Delta_{n,\mathbf{K}}^{(R)}(D)$ are summarized in Fig. 6(a,c) inset, both are found to

be in exponential forms

$$t_{n,\mathbf{K}}^{(R)}(D) = t_{n,\mathbf{K}}^{(R)}(D_m)e^{-(D-D_m)/\delta D_{n,\mathbf{K}}^{(T)}},$$
$$\Delta_{n,\mathbf{K}}^{(R)}(D) = \Delta_{n,\mathbf{K}}^{(R)}(D_m)e^{-(D-D_m)/\delta D_{n,\mathbf{K}}^{(\Delta)}}. \quad (9)$$

The fitting parameters $\delta D_{n,\mathbf{K}}^{(T)}$ and $\delta D_{n,\mathbf{K}}^{(\Delta)}$ are given in Table V.

In long-wavelength moiré patterns of $M\mathrm{Si}_2X_4$, different local regions in a moiré supercell can be characterized by their spatially-varying interlayer translations $\mathbf{r}_0$. The resultant local stacking registries at different positions thus have different interlayer separations, similar to bilayer TMDs which has been observed in experiments [75]. We expect the local interlayer separation to be close to $D_{eq}(\mathbf{r}_0)$. After taking into account the dependence of the equilibrium interlayer distance $D_{eq}(\mathbf{r}_0)$ on $\mathbf{r}_0$, the splitting between the two sub-bands becomes

$$\Delta\varepsilon_{n,\mathbf{k}}^{(R,eq)}(\mathbf{r}_0) = 2\sqrt{\left|\delta E_{n,\mathbf{k}}^{(R)}(D_{eq}(\mathbf{r}_0),\mathbf{r}_0)\right|^2 + \left|T_{n,\mathbf{k}}^{(R)}(D_{eq}(\mathbf{r}_0),\mathbf{r}_0)\right|^2}. \quad (10)$$

The numerically calculated values of $\Delta\varepsilon_{n,\mathbf{k}}^{(R,eq)}(\mathbf{r}_0)$ and the corresponding analytic forms from Eq. (10) are shown in Fig. 6(b,d) as symbols and curves, respectively.

The two layers of H-type homobilayers are always related by the inversion symmetry, resulting in $\delta E_{n,\mathbf{k}}^{(H)}(D,\mathbf{r}_0) = 0$ for any wave vector $\mathbf{k}$. When the SOC effect is not considered, the splitting $\Delta\varepsilon_{n,\mathbf{K}}^{(H)}(D,\mathbf{r}_0) = 2|T_{n,\mathbf{K}}^{(H)}(D,\mathbf{r}_0)|$ between the two sub-bands directly reflects the interlayer hopping strengths at $\mathbf{K}$. Fig. 7(a,b) show numerical values of $\Delta\varepsilon_{v,\mathbf{K}}^{(H)}(D,\mathbf{r}_0)$ as functions of $\mathbf{r}_0$ in H-type homobilayer $M\mathrm{Si}_2X_4$ under four different $D$ values, and the corresponding fitting curves using Eq. (7). Unlike the R-type case where keeping only the main hopping term $t_{n,\mathbf{K}}^{(R)}(D)$ can already leads to very good fittings, in H-type homobilayer $M\mathrm{Si}_2\mathrm{P}_4$ and $M\mathrm{Si}_2\mathrm{As}_4$, both the main term $t_{v,\mathbf{K}}^{(H)}(D)$ and higher-order terms $t_{v,\mathbf{K}}'^{(H)}(D)$, $t_{v,\mathbf{K}}''^{(H)}(D)$ need to be included to ensure good fitting results, especially under small values of $D$. The fitting results are again well described by $t_{v,\mathbf{K}}^{(H)}(D) = t_{v,\mathbf{K}}^{(H)}(D_m)e^{-(D-D_m)/\delta D_{v,\mathbf{K}}^{(T)}}$, $t_{v,\mathbf{K}}'^{(H)}(D) = t_{v,\mathbf{K}}'^{(H)}(D_m)e^{-(D-D_m)/\delta D_{v,\mathbf{K}}'^{(T)}}$ and $t_{v,\mathbf{K}}''^{(H)}(D) = t_{v,\mathbf{K}}''^{(H)}(D_m)e^{-(D-D_m)/\delta D_{v,\mathbf{K}}''^{(T)}}$, with $t_{v,\mathbf{K}}^{(H)}(D_m)$, $t_{v,\mathbf{K}}'^{(H)}(D_m)$ and $t_{v,\mathbf{K}}''^{(H)}(D_m)$ summarized in Table IV and $\delta D_{v,\mathbf{K}}^{(T)}$, $\delta D_{v,\mathbf{K}}'^{(T)}$ and $\delta D_{v,\mathbf{K}}''^{(T)}$ given in Table V. The values in Table V indicate that $\delta D_{v,\mathbf{K}}'^{(T)}$ and $\delta D_{v,\mathbf{K}}''^{(T)}$ of higher-order terms are significantly smaller than $\delta D_{v,\mathbf{K}}^{(T)}$ of the main term. The sub-band splitting $\Delta\varepsilon_{v,\mathbf{k}}^{(H,eq)}(\mathbf{r}_0) = 2\left|T_{v,\mathbf{k}}^{(H)}(D_{eq}(\mathbf{r}_0),\mathbf{r}_0)\right|$ under the equilibrium interlayer distance $D_{eq}(\mathbf{r}_0)$ are shown in Fig. 7(c). Note that $\Delta\varepsilon_{c,\mathbf{K}}^{(H)}(D,\mathbf{r}_0)$ of the conduction band is found to be rather close to zero thus not shown, which indicates that the $\mathbf{K}$-point conduction band interlayer hopping in H-type homobilayer

$M$Si$_2$X$_4$ is negligible.

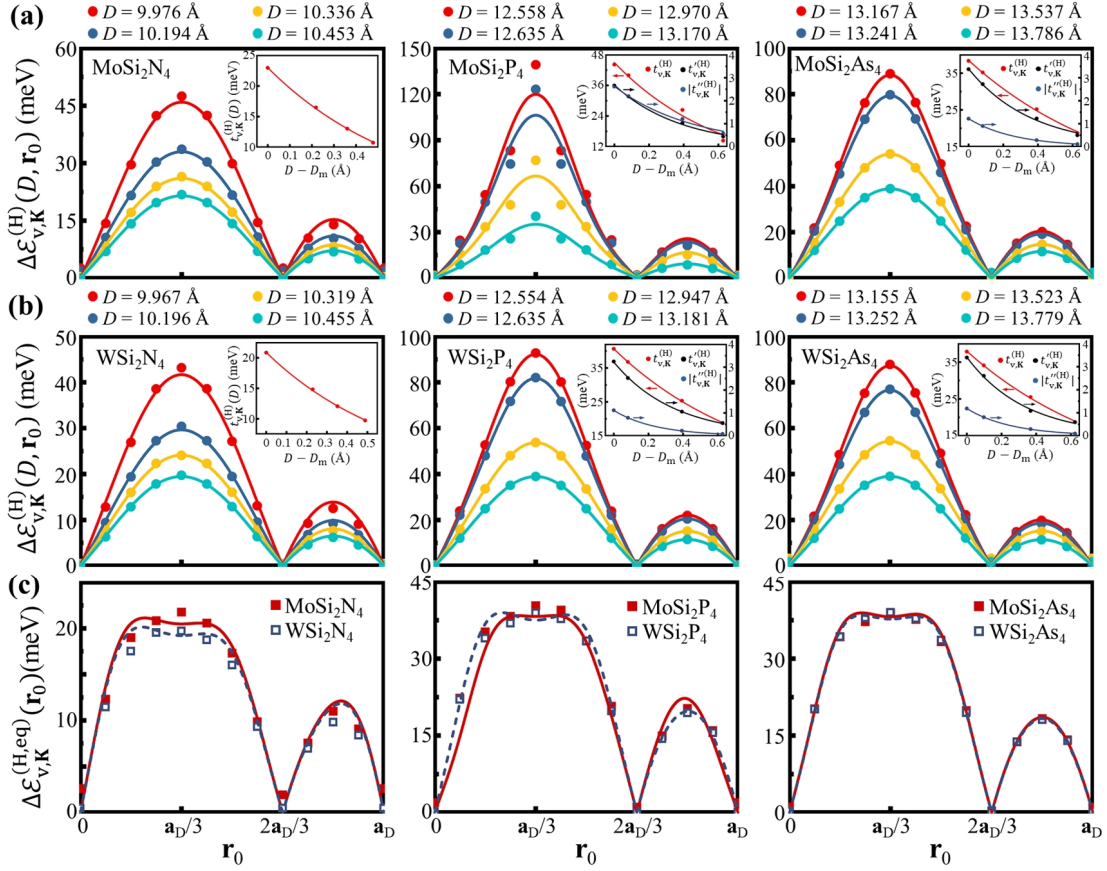

Fig. 7 (a) The **K**-point energy splitting $\Delta\mathcal{E}_{v,\mathbf{K}}^{(H)}$ between the two valence sub-bands in H-type homobilayer MoSi$_2$X$_4$. (b) $\Delta\mathcal{E}_{v,\mathbf{K}}^{(H)}$ in H-type homobilayer WSi$_2$X$_4$. (c) The splitting $\Delta\mathcal{E}_{v,\mathbf{k}}^{(H,eq)}(\mathbf{r}_0)$ after taking into account the dependence of the equilibrium interlayer distance $D_{eq}(\mathbf{r}_0)$ on $\mathbf{r}_0$. In (a-c), symbols correspond to the numerically calculated values, and curves are analytical fittings.

Table V. The decay length obtained by fitting the curves in Fig. 6 and Fig. 7 insets. The interlayer hopping at $\mathbf{K}_c$ for R-type homobilayer MoSi$_2$P$_4$ and MoSi$_2$As$_4$ is close to zero, thus the corresponding decay length $\delta D_{c,\mathbf{K}}^{(T)}$ has a large error and is not shown.

|  |  | MoSi$_2$N$_4$ | WSi$_2$N$_4$ | MoSi$_2$P$_4$ | WSi$_2$P$_4$ | MoSi$_2$As$_4$ | WSi$_2$As$_4$ |
|---|---|---|---|---|---|---|---|
| R-type | $\delta D_{c,\mathbf{K}}^{(T)}$ (Å) | 0.118 | 0.103 | — | 0.630 | — | 0.685 |
|  | $\delta D_{c,\mathbf{K}}^{(\Delta)}$ (Å) | 0.398 | 0.349 | 0.176 | 0.406 | 0.370 | 0.508 |
|  | $\delta D_{v,\mathbf{K}}^{(T)}$ (Å) | 0.581 | 0.538 | 0.751 | 0.773 | 0.797 | 0.826 |
|  | $\delta D_{v,\mathbf{K}}^{(\Delta)}$ (Å) | 0.392 | 0.352 | 0.181 | 0.479 | 0.387 | 0.548 |
| H-type | $\delta D_{v,\mathbf{K}}^{(T)}$ (Å) | 0.635 | 0.646 | 0.630 | 0.818 | 0.892 | 0.899 |
|  | $\delta D_{v,\mathbf{K}}'^{(T)}$ (Å) | — | — | 0.384 | 0.351 | 0.362 | 0.338 |
|  | $\delta D_{v,\mathbf{K}}''^{(T)}$ (Å) | — | — | 0.459 | 0.240 | 0.264 | 0.257 |

**(b) Interlayer couplings at $\Gamma_v$ and layer-hybridizations**

In homobilayer MoSi$_2$N$_4$ and WSi$_2$N$_4$, the valence band edge is located at $\Gamma_v$ (see Fig. 4(a)), whose sub-band energies are described by

$$\mathcal{E}_{v\pm,\Gamma}^{(R)}(D,\mathbf{r}_0) = E_{v,\Gamma}^{(R)} + \delta_{v,\Gamma}^{(R)}(D)[|f_+(\mathbf{r}_0)|^2 + |f_-(\mathbf{r}_0)|^2]$$
$$\pm \sqrt{\left(\Delta_{v,\Gamma}^{(R)}(D)[|f_+(\mathbf{r}_0)|^2 - |f_-(\mathbf{r}_0)|^2]\right)^2 + \left|t_{v,\Gamma}^{(R)}(D)\right|^2}, \quad (11)$$
$$\mathcal{E}_{v\pm,\Gamma}^{(H)}(D,\mathbf{r}_0) = E_{v,\Gamma}^{(H)} + \delta_{v,\Gamma}^{(H)}(D)|f_+(\mathbf{r}_0)|^2 + \Delta_{v,\Gamma}^{(H)}(D)|f_-(\mathbf{r}_0)|^2 \pm \left|t_{v,\Gamma}^{(H)}(D)\right|.$$

Note that the interlayer hopping $T_{v,\Gamma}^{(R/H)}(D,\mathbf{r}_0) \approx t_{v,\Gamma}^{(R/H)}(D)$ at $\Gamma$ becomes independent on $\mathbf{r}_0$, which is distinct from $T_{n,K}^{(R/H)}(D,\mathbf{r}_0)$ at $\mathbf{K}$. The variation of $\mathcal{E}_{v\pm,\Gamma}^{(R)}(D,\mathbf{r}_0)$ for different $\mathbf{r}_0$ comes from the $\mathbf{r}_0$-dependent diagonal energy shifts of the basis Bloch states in two decoupled layers.

In Fig. 8(a), the symbols correspond to our calculated values of $\mathcal{E}_{v\pm,\Gamma}^{(R)}(D_m,\mathbf{r}_0)$ for R-type homobilayer MoSi$_2$N$_4$ and WSi$_2$N$_4$ at their minimum equilibrium interlayer distances $D_m$, and the curves are fittings using Eq. (11). The symbols and curves show perfect agreement, with fitting parameters summarized in Table VI. In Fig. 8(b), we show the energy splitting $\Delta\mathcal{E}_{v,\Gamma}^{(R)}(D,\mathbf{r}_0)$ at four different values of $D$. The obtained values of $\Delta_{v,\Gamma}^{(R)}(D)$ and $t_{v,\Gamma}^{(R)}(D)$ decay exponentially with $D$, which can be written as $\Delta_{v,\Gamma}^{(R)}(D) = \Delta_{v,\Gamma}^{(R)}(D_m)e^{-(D-D_m)/\delta D_{v,\Gamma}^{(\Delta)}}$ and $t_{v,\Gamma}^{(R)}(D) = t_{v,\Gamma}^{(R)}(D_m)e^{-(D-D_m)/\delta D_{v,\Gamma}^{(T)}}$ with $\delta D_{v,\Gamma}^{(\Delta)}$ and $\delta D_{v,\Gamma}^{(T)}$ summarized in Table VI. A comparison with the $\mathbf{K}$-point fitting results in Table IV/V indicates that $\Delta_{v,\Gamma}^{(R)}(D_m) \approx \Delta_{c,K}^{(R)}(D_m) \approx \Delta_{v,K}^{(R)}(D_m) \gg \delta_{v,\Gamma}^{(R)}(D_m), \delta_{v,K}^{(R)}(D_m)$ and $\delta D_{v,\Gamma}^{(\Delta)} \approx \delta D_{v,K}^{(\Delta)} \approx \delta D_{c,K}^{(\Delta)}$, again implying that the diagonal energy shifts of the basis Bloch states are induced predominantly by the $\mathbf{k}$-independent spontaneous electrical polarization across the bilayer interface.

In H-type homobilayer MoSi$_2$N$_4$ and WSi$_2$N$_4$, the calculated energy splitting $\Delta\mathcal{E}_{v,\Gamma}^{(H)}(D_m,\mathbf{r}_0)$ is also found to be independent on $\mathbf{r}_0$, in agreement with our expectation. The resultant $t_{v,\Gamma}^{(H)}(D_m) \approx 46$ meV for the H-type MoSi$_2$N$_4$ and WSi$_2$N$_4$ is close to $t_{v,\Gamma}^{(R)}(D_m)$ of the R-type. Such values are larger but in the same order as $t_{v,K}^{(R/H)}(D_m)$. In contrast, the interlayer hopping at $\Gamma_v$ in bilayer TMDs is as large as several hundred meV, one order of magnitude stronger [68].

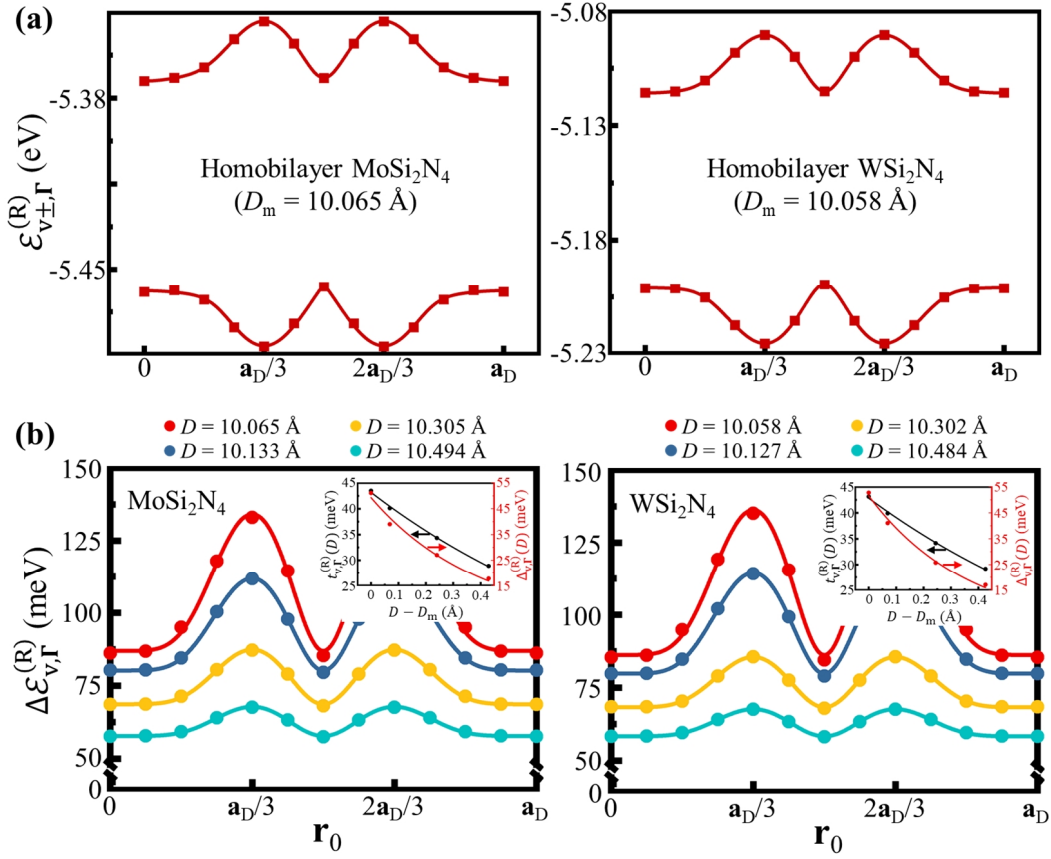

Fig. 8 (a) The two valence sub-band energies $\mathcal{E}_{v\pm,\Gamma}^{(R)}$ at $\Gamma$ in R-type homobilayer $MoSi_2N_4$ and $WSi_2N_4$ at their minimum equilibrium interlayer distances $D_m$. (b) The sub-band splitting $\Delta\mathcal{E}_{v,\Gamma}^{(R)}$ in R-type homobilayer $MoSi_2N_4$ and $WSi_2X_4$ under different interlayer distances $D$. The symbols in insets show fitting parameters $t_{v,\Gamma}^{(R)}(D)$ and $\Delta_{v,\Gamma}^{(R)}(D)$ as functions of $D$, both can be well fit by exponential decays (curves).

Table VI. Fitting parameters for the curves in Fig. 8 using Eq. (4).

|  | $\delta_{v,\Gamma}^{(R)}(D_m)$ | $\Delta_{v,\Gamma}^{(R)}(D_m)$ | $\delta D_{v,\Gamma}^{(\Delta)}$ | $t_{v,\Gamma}^{(R)}(D_m)$ | $\delta D_{v,\Gamma}^{(T)}$ | $t_{v,\Gamma}^{(H)}(D_m)$ |
|---|---|---|---|---|---|---|
| $MoSi_2N_4$ | -0.86 meV | 51.08 meV | 0.398 Å | 43.48 meV | 1.057 Å | 45.9 meV |
| $WSi_2N_4$ | -0.55 meV | 52.75 meV | 0.363 Å | 43.12 meV | 1.095 Å | 45.5 meV |

Fig. 9(a) schematically illustrates the sub-bands near the corresponding band extrema $K_{c1}$, $K_{c2}$, $K_{v1}$, $K_{v2}$ and $\Gamma_{v1}$, $\Gamma_{v2}$ of the homobilayer, which can be layer-hybridized due to the interlayer hopping. For H-type configurations without including SOC, the two layers are related by an inversion thus each state is an equal superposition of the two layers. However when the SOC is included, the interlayer hopping at $K$ in H-type homobilayers is suppressed by the strong SOC splitting (0.1 to 0.4 eV at $K_v$ and 3 to 26 meV at $K_c$), resulting in its negligible layer hybridization.

On the other hand, in R-type homobilayers the layer-hybridization is determined by the competition between the interlayer hopping $T_{n,\mathbf{k}}^{(R)}(D,\mathbf{r}_0)$ and the diagonal energy difference $\delta E_{n,\mathbf{k}}^{(R)}(D,\mathbf{r}_0)$ between basis Bloch states. Since $\delta E_{n,\mathbf{k}}^{(R)}(D,\mathbf{r}_0)$ is nearly independent on $\mathbf{k}$, the difference between interlayer couplings at $\Gamma$ and $\mathbf{K}$ comes from the fact that $T_{n,\mathbf{K}}^{(R)}(D,\mathbf{r}_0) \approx t_{n,\mathbf{K}}^{(R)}(D)f_0(\mathbf{r}_0)$ varies sensitively with $\mathbf{r}_0$ whereas $T_{v,\Gamma}^{(R)}(D,\mathbf{r}_0) \approx t_{v,\Gamma}^{(R)}(D)$ is independent on $\mathbf{r}_0$. The largest difference between $\Gamma$ and $\mathbf{K}$ occurs for the AB (or BA) stacking, where $T_{n,\mathbf{K}}^{(R)}\left(D,\frac{\mathbf{a}_D}{3}\right) = 0$ but $T_{v,\Gamma}^{(R)}\left(D,\frac{\mathbf{a}_D}{3}\right) \neq 0$ thus the two sub-bands have finite (zero) layer-hybridizations at $\Gamma$ ($\mathbf{K}$). Here we use the layer-polarization $\langle\hat{\sigma}_z\rangle \equiv \langle\psi|\hat{\sigma}_z|\psi\rangle$ to quantify the layer-hybridization of a given state $|\psi\rangle$, such that $\langle\hat{\sigma}_z\rangle \approx 0$ ($\pm 1$) corresponds to the largest (a vanishing) layer-hybridization. Fig. 9(b) shows our calculated $\langle\hat{\sigma}_z\rangle$ at $\Gamma_{v1}$ and $\Gamma_{v2}$ as functions of $D$ in AB-stacked homobilayer $MoSi_2N_4$ and $WSi_2N_4$. At the equilibrium interlayer distance which corresponds to the minimum value $D_m$, $\Gamma_{v1}$ and $\Gamma_{v2}$ show weak layer-hybridizations with layer-polarizations reaching $\pm 0.8$. This is further confirmed by the calculated charge density iso-surfaces shown in Fig. 9(c). The iso-surfaces at a value of 0.02 e·Å$^{-3}$ are fully located in the upper-layer (lower-layer) of AB-stacked homobilayer $MoSi_2N_4$ and $WSi_2N_4$ for $\Gamma_{v1}$, $\mathbf{K}_{v1}$ and $\mathbf{K}_{c1}$ ($\Gamma_{v2}$, $\mathbf{K}_{v2}$ and $\mathbf{K}_{c2}$). These are in sharp contrast to those in bilayer TMDs, where the layer-polarization at $\Gamma_v$ is near 0 due to the strong interlayer hopping strength (several hundred meV) [68]. The charge density iso-surfaces of AB-stacked homobilayer $MoSi_2P_4$ and $MoSi_2As_4$ are also shown in Fig. 9(c), where $\Gamma_{v1}$ and $\Gamma_{v2}$ become strongly layer-hybridized. This can be attributed to their much stronger interlayer hopping at $\Gamma_v$ originating from the ~ 8% $p_z$-orbital fraction of the outer X atoms (in contrast to ~ 4% in $MoSi_2N_4$ and $WSi_2N_4$), see Table II.

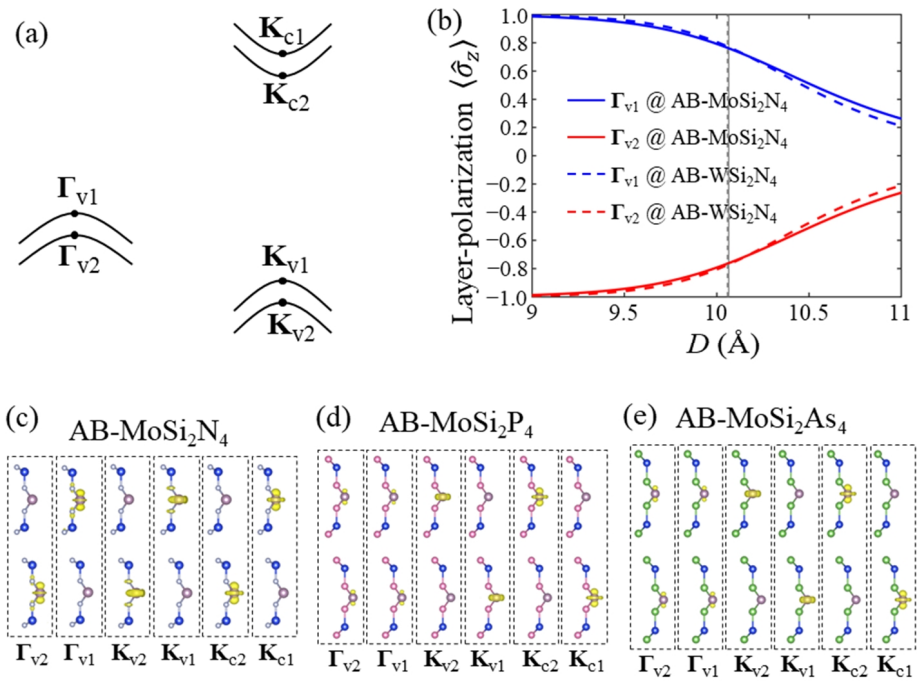

Fig. 9 (a) A schematic illustration of the sub-bands near the corresponding band extrema $K_{c1}$, $K_{c2}$, $K_{v1}$, $K_{v2}$ and $\Gamma_{v1}$, $\Gamma_{v2}$ in homobilayers. (b) The calculated layer-polarizations at $\Gamma_{v1}$ and $\Gamma_{v2}$ as functions of $D$ in AB-stacked homobilayer $MoSi_2N_4$ and $WSi_2N_4$. The vertical solid (dashed) gray line corresponds to the minimum interlayer distance $D_m$ of $MoSi_2N_4$ ($WSi_2N_4$). (c-e) The calculated charge density iso-surfaces at 0.02 e·Å$^{-3}$ for the band extrema $K_{c1}$, $K_{c2}$, $K_{v1}$, $K_{v2}$ and $\Gamma_{v1}$, $\Gamma_{v2}$ of AB-stacked homobilayer (c) $MoSi_2N_4$, (d) $MoSi_2P_4$ and (e) $MoSi_2As_4$.

### (c) Honeycomb lattice models for holes at $\Gamma_v$ and $K_v$ in moiré patterned homobilayers

By introducing a small twist angle between the two layers, a long wavelength moiré superlattice pattern can form in homobilayer $MSi_2X_4$. For a local region with a size much smaller than the moiré supercell, its local atomic registry is indistinguishable from an R- or H-type commensurate bilayer with a spatially varying interlayer translation $r_0(\mathbf{r})$. Here $\mathbf{r}$ is the center position of the local region. Below we focus on near R-type moiré patterns of $MoSi_2N_4$ or $WSi_2N_4$ with interlayer twist angles close to 0, where local regions with AA, AB and BA stacking registries appear at different positions in the moiré supercell (see Fig. 10(a) for an illustration). The local electronic structure can then be approximated by that of the commensurate homobilayer under the interlayer translation $r_0(\mathbf{r})$ and equilibrium interlayer distance $D_{eq}(r_0(\mathbf{r}))$.

Using a low-energy continuum model, one can write the Hamiltonian near a valence band maximum ($K_v$- or $\Gamma_v$-valley) as [62,63]

$$\hat{H}_{v,\text{moiré}} \approx E_{v,K/\Gamma}^{(R)} - \frac{\hbar^2}{2m^*}\frac{\partial^2}{\partial \mathbf{r}^2} + \left(\Delta_{\text{moiré}}(\mathbf{r}) + \frac{V}{2}\right)\hat{\sigma}_z + T_{\text{moiré}}(\mathbf{r})\hat{\sigma}_+ + T_{\text{moiré}}^*(\mathbf{r})\hat{\sigma}_-. \quad (12)$$

On the above right-hand-side, the first term is treated as a constant energy, where we have ignored the layer-independent modulation term $\delta_{n,K/\Gamma}^{(R)}(D)$ in Eq. (6) since it is rather weak in $MoSi_2N_4$ and $WSi_2N_4$; the second term is the kinetic energy with $m^*$ the effective mass which accounts for the band dispersion; the last three terms describe the position-dependent layer-pseudospin $(\hat{\sigma}_x, \hat{\sigma}_y, \hat{\sigma}_z)$, with $V$ an externally applied interlayer bias. In $K_v$-valley, $\Delta_{\text{moiré}}(\mathbf{r})$ and $T_{\text{moiré}}(\mathbf{r})$ have the following forms

$$\Delta_{\text{moiré}}(\mathbf{r}) = \Delta_{v,K}^{(R)}(D_m)e^{-\alpha_K|f_0(r_0(\mathbf{r}))|^2}\left[|f_+(r_0(\mathbf{r}))|^2 - |f_-(r_0(\mathbf{r}))|^2\right],$$
$$T_{\text{moiré}}(\mathbf{r}) = t_{v,K}^{(R)}(D_m)e^{-\beta_K|f_0(r_0(\mathbf{r}))|^2}e^{-i\mathbf{K}_m\cdot\mathbf{r}}f_0(r_0(\mathbf{r})), \quad (13)$$

with $\alpha_K = \Delta D_0/\delta D_{v,K}^{(\Delta)}$, $\beta_K = \Delta D_0/\delta D_{v,K}^{(T)}$ and $\mathbf{K}_m$ the moiré Brillouin zone corner (see Fig. 10(a) inset). In $\Gamma_v$-valley, the forms of $\Delta_{\text{moiré}}(\mathbf{r})$ and $T_{\text{moiré}}(\mathbf{r})$ are

$$\Delta_{\text{moiré}}(\mathbf{r}) = \Delta_{v,\Gamma}^{(R)}(D_m)e^{-\alpha_\Gamma|f_0(r_0(\mathbf{r}))|^2}\left[|f_+(r_0(\mathbf{r}))|^2 - |f_-(r_0(\mathbf{r}))|^2\right],$$
$$T_{\text{moiré}}(\mathbf{r}) = t_{v,\Gamma}^{(R)}(D_m)e^{-\beta_\Gamma|f_0(r_0(\mathbf{r}))|^2}, \quad (14)$$

with $\alpha_\Gamma = \Delta D_0/\delta D_{v,\Gamma}^{(\Delta)}$, $\beta_\Gamma = \Delta D_0/\delta D_{v,\Gamma}^{(T)}$.

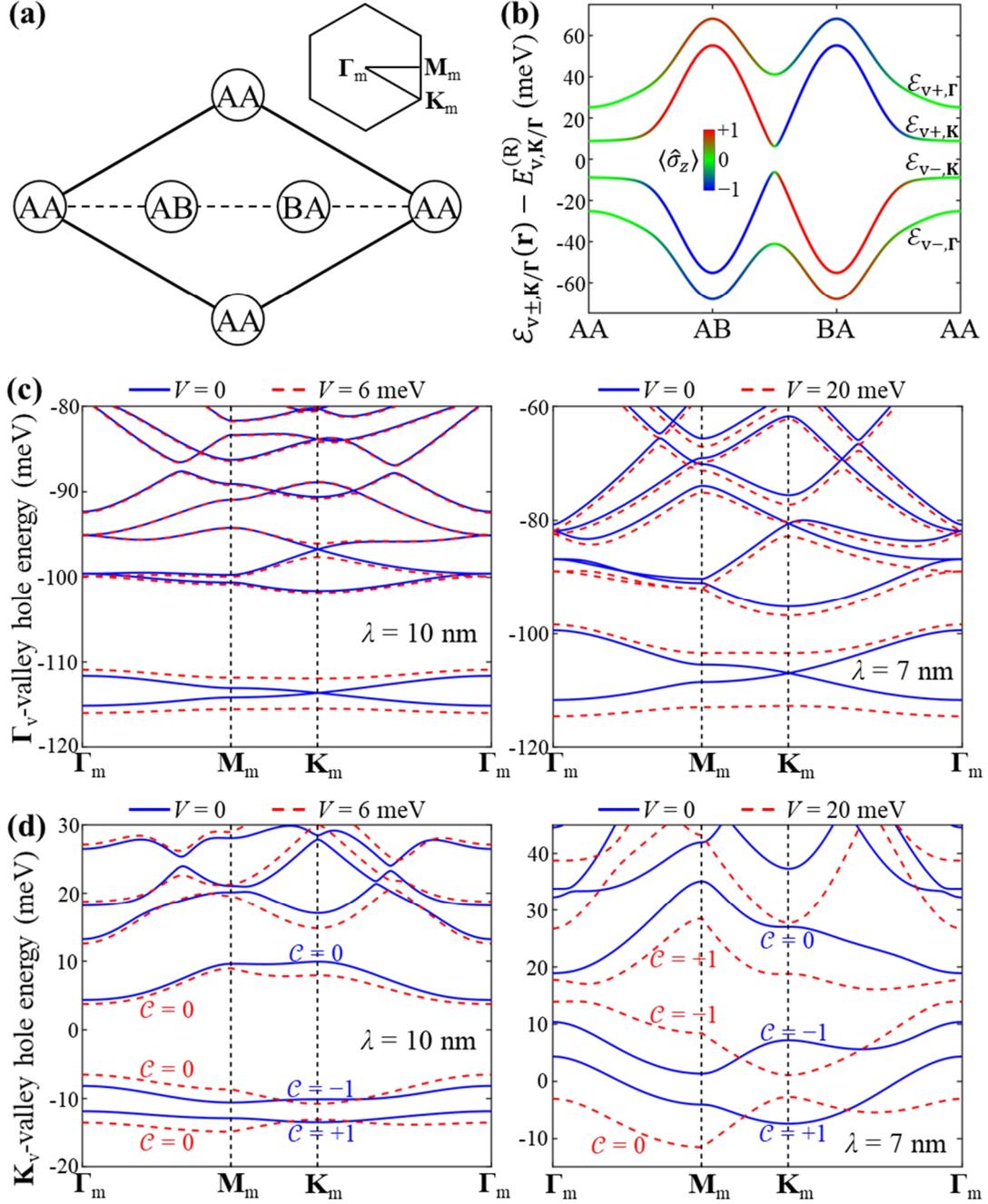

Fig. 10 (a) A schematic illustration of a supercell of the near R-type homobilayer $M$Si$_2X_4$ moiré pattern. Local regions with AA, AB and BA stacking registries appear at different positions along the long diagonal line (dashed line). The inset shows the moiré mini Brillouin zone. (b) The layer-pseudospin eigen-energies $\mathcal{E}_{v,\pm}(\mathbf{r})$ and the corresponding layer-polarizations of valence electrons in $\mathbf{\Gamma}_v$ and $\mathbf{K}_v$ valleys as function of $\mathbf{r}$ along the long diagonal line in (a), under zero interlayer bias. (c) The moiré mini bands for the $\mathbf{\Gamma}_v$-valley hole of homobilayer WSi$_2$N$_4$ under wavelengths of $\lambda = 10$ nm (left panel) and 7 nm (right panel). Blue solid (red dashed) lines corresponding those with $V = 0$ ($V \neq 0$). (d) The moiré mini bands for the $\mathbf{K}_v$-valley hole. Topological Chern numbers $\mathcal{C}$ of the three lowest-energy mini bands are also given.

We consider a homobilayer WSi$_2$N$_4$ moiré pattern with an interlayer twist angle $\theta$, where the upper-layer (lower-layer) Brillouin zone corners are given by $\mathbf{K}$ ($\hat{C}_\theta \mathbf{K}$ with $\hat{C}_\theta$ the $\theta$-angle rotation operation) and its $\pi/3$ rotations, then $\mathbf{K}_m \equiv \mathbf{K} - \hat{C}_\theta \mathbf{K}$. The corresponding effective mass is $m^* \approx -\mathbf{0.5}m_0$ ($m^* \approx -\mathbf{1.8}m_0$) near $\mathbf{K}_v$ ($\mathbf{\Gamma}_v$), with $m_0$ the free electron mass. When the two layers are treated as rigid and the lattice reconstruction effect is ignored, $\mathbf{r}_0(\mathbf{r})$ is a linear function of $\mathbf{r}$ which satisfies $\mathbf{K} \cdot \mathbf{r}_0(\mathbf{r}) = \mathbf{K}_m \cdot \mathbf{r}$ (mod $2\pi$) [34]. By diagonalizing the 2×2 layer-pseudospin matrix in Eq. (12), we get the layer-pseudospin eigen-energies $\mathcal{E}_{v\pm,K/\Gamma}(\mathbf{r})$ and the corresponding layer-polarizations $\langle \hat{\sigma}_z \rangle$ as functions of $\mathbf{r}$ (see Fig. 10(b)). Note that $-\mathcal{E}_{v+,K/\Gamma}(\mathbf{r})$ can be viewed as the potential energy experienced by the low-energy hole. At a first sight, it seems that $\mathbf{K}_v$- and $\mathbf{\Gamma}_v$-valley holes have rather similar potential profiles and layer-polarizations: both potential profiles form honeycomb patterns with potential minima located at AB and BA sites, also the corresponding layer-polarizations vary with position which cross from $\approx +1$ at AB to $\approx -1$ at BA. However, the interlayer hopping $T_{\text{moiré}}(\mathbf{r})$, which determines the in-plane layer-pseudospin $(\langle \hat{\sigma}_x \rangle, \langle \hat{\sigma}_y \rangle)$, has different forms for $\mathbf{K}_v$- and $\mathbf{\Gamma}_v$-valley holes. As a result, these two cases have qualitatively different band energies and topologies as discussed below.

$T_{\text{moiré}}(\mathbf{r})$ has a constant phase for the $\mathbf{\Gamma}_v$-valley hole. The corresponding layer-pseudospin in-plane direction does not change with position, resulting in a trivial band topology similar to the gapless or gapped graphene. Fig. 10(c) shows the calculated moiré mini bands for the $\mathbf{\Gamma}_v$-valley hole under wavelengths of $\lambda = 10$ nm (left panel) and 7 nm (right panel), with the blue solid (red dashed) lines obtained under an interlayer bias $V = 0$ ($V \neq 0$). The two lowest-energy bands can be described by a graphene model with the two sublattice sites being the $s$-type wave packets localized at AB and BA. These two wave packets are degenerate under $V = 0$, resulting in a low-energy massless Dirac cone at $\mathbf{K}_m$. A finite value of $V$ breaks the sublattice degeneracy, and the two lowest-energy bands become well separated in energy which have finite $\mathbf{k}$-dependent Berry curvatures. The topological Chern numbers $\mathcal{C}$, however, are zero for both bands. Note that the 4th and 5th bands also cross at $\mathbf{K}_m$, forming a higher-energy Dirac cone whose main compositions are the $p$-type wave packets localized at AB and BA.

Similar to homobilayer TMDs [62,63], the spatially varying phase of $T_{\text{moiré}}(\mathbf{r})$ for the $\mathbf{K}_v$-valley hole can give rise to its non-trivial band topologies. Fig. 10(d) shows the calculated moiré mini bands of the $\mathbf{K}_v$-valley hole. Unlike the $\mathbf{\Gamma}_v$-valley case, the two lowest-energy bands are well separated in energy when $V = 0$. The Chern numbers of the three lowest-energy bands under $V = 0$ are $\mathcal{C} = +1, -1$ and 0, indicating non-trivial band topologies. Note that $-\mathbf{K}_v$-valley is simply the time reversal of $\mathbf{K}_v$-valley with opposite Chern numbers. When the valence band edge is shifted from $\mathbf{\Gamma}_v$ to $\mathbf{K}_v$ by a strain [49,53], homobilayer $WSi_2N_4$ moiré patterns can serve as a platform for realizing the quantum spin/valley Hall insulator. Meanwhile, the band energies and Chern numbers are rather sensitive to the values of wavelength $\lambda$ and interlayer bias $V$. Under $\lambda = 10$ nm, applying a bias value of $V = 6$ meV can change the Chern numbers to $\mathcal{C} = 0$,

0 and 0. Whereas for a relatively short wavelength $\lambda = 7$ nm, $V = 20$ meV can change the Chern numbers to $\mathcal{C} = 0$, $-1$ and $+1$, see the right panel of Fig. 10(d). These results indicate that both band energies and topologies are highly tunable through varying the moiré wavelength and interlayer bias.

## IV. Conclusion

In conclusion, using the combination of first-principles calculations and analytical investigations, we have systematically investigated the interlayer coupling effect in $\mathbf{K}_{c/v}$- and $\mathbf{\Gamma}_v$-valleys of commensurate $M\text{Si}_2X_4$ homobilayers under various stacking registries. The equation forms of equilibrium interlayer distances, layer energy differences and interlayer hopping strengths as functions of the interlayer distance and interlayer translation are obtained, with the important parameters determined from fitting the numerical results. Although the equilibrium interlayer distance of bilayer $M\text{Si}_2X_4$ (~ 10 Å) is significantly larger than that of bilayer TMDs (~ 6 Å), we find that their interlayer coupling strengths at $\mathbf{K}_c$ and $\mathbf{K}_v$ are actually in the same order. In homobilayer $\text{MoSi}_2\text{N}_4$ and $\text{WSi}_2\text{N}_4$, the layer energy difference from the interlayer charge redistribution can reach 0.1 eV under AB/BA stacking, in the same order to those recently observed in bilayer hexagonal boron nitride and TMDs. The interlayer hopping strengths in all six $M\text{Si}_2X_4$ materials are all as large as several tens meV at $\mathbf{K}_v$ and $\mathbf{\Gamma}_v$, and ~ 1 meV at the conduction band edge $\mathbf{K}_c$. In near R-type moiré patterned homobilayer $M\text{Si}_2X_4$, the spatially varying interlayer coupling effect can introduce a moiré potential with a modulation range ~ 50 meV to the conduction/valence band edge carrier. In near R-type moiré patterns of homobilayer $M\text{Si}_2\text{N}_4$, the moiré potential has a modulation range ~ 50 meV which realizes a honeycomb lattice model for low-energy carriers. For $\mathbf{\Gamma}_v$-valley holes in homobilayer $M\text{Si}_2\text{N}_4$, the moiré pattern can realize a gapless or gapped graphene model. Meanwhile, for $\mathbf{K}_v$-valley holes in homobilayer $M\text{Si}_2\text{N}_4$, the moiré pattern can realize a honeycomb lattice model with non-trivial band topologies, where the band energies and Chern numbers can both be tuned by applying an interlayer bias. Our findings can provide a guidance for designing electronic and optoelectronic applications based on $M\text{Si}_2X_4$ bilayer structures.

**Acknowledgement.** H.Y. acknowledges support by NSFC under grant No. 12274477, and the Department of Science and Technology of Guangdong Province in China (2019QN01X061).

## Appendix: Computation methods

Our calculations are performed based on DFT [76]. The Vienna Ab initio Simulation Package (VASP) [76,77] was adopted for geometry optimization and electronic properties. The plane-wave cutoff energy is set to 600 eV. The residual forces have converged to less than 0.001 eV/Å and the total energy difference to less than $10^{-6}$ eV. A $\Gamma$-centered Brillouin zone $k$-point sampling grid using Monkhorst-Pack scheme [78] of $21 \times 21 \times 1$ mesh was used for the structural optimization and a grid of $42 \times 42 \times 1$ mesh for the property calculations. The electron-ion interactions are modeled using the projector augmented wave (PAW) potentials [79,80]. The generalized gradient approximation (GGA) with the Perdew-Burke-Ernzerhof (PBE) [81] functional was chosen for the exchange-correlation interactions in the calculations. Considering van der Waals corrections in bilayers, we used DFT-D3 [82] method with the Grimme scheme in the calculations. A vacuum of 15 Å is used in all the calculations to avoid interaction between the neighboring slabs from periodic. The consideration of SOC is limited to the calculation of electronic band structures in $M$Si$_2$X$_4$ monolayers.

The band structures are further checked by the more accurate Heyd-Scuseria-Ernzerhof hybrid functional method (HSE06) [83], which leads to different bandgap values shown in Fig. A1(a) and Table A1. Note that the interlayer coupling effect, which is encoded in the energy splitting between the two conduction/valence sub-bands of homobilayer $M$Si$_2$X$_4$, is barely affected by the functional method we use in the calculation (see Fig. A1(b,c) and Table A1). In this work, we analyze the interlayer coupling based on PBE band structures.

Table A1. A comparison of key band features obtained from DFT calculations (SOC not included) using PBE and HSE06 functional methods, for MoSi$_2$N$_4$ bilayer under high-symmetry stacking registries at the equilibrium interlayer distance $D_{eq}$. $E_g$ is the global bandgap, $\Delta\varepsilon_{c,K}(D_{eq})$, $\Delta\varepsilon_{v,K}(D_{eq})$, and $\Delta\varepsilon_{v,\Gamma}(D_{eq})$ correspond to energy splittings between the two sub-bands at the band extrema $\mathbf{K}_c$, $\mathbf{K}_v$ and $\mathbf{\Gamma}_v$, respectively.

| Stacking | Method | $E_g$ (eV) | $\Delta\varepsilon_{c,K}(D_{eq})$ (meV) | $\Delta\varepsilon_{v,K}(D_{eq})$ (meV) | $\Delta\varepsilon_{v,\Gamma}(D_{eq})$ (meV) |
|---|---|---|---|---|---|
| AB′ | PBE | 1.769 | 2.607 | 2.586 | 92.929 |
|  | HSE06 | 2.355 | 0.261 | 0.457 | 88.601 |
| AA′ | PBE | 1.811 | 0.009 | 21.762 | 59.392 |
|  | HSE06 | 2.388 | 0.265 | 19.512 | 56.018 |
| A′B | PBE | 1.797 | 1.767 | 1.915 | 75.325 |
|  | HSE06 | 2.376 | 0.347 | 1.349 | 71.281 |
| AA | PBE | 1.807 | 4.106 | 20.712 | 57.613 |
|  | HSE06 | 2.386 | 0.301 | 18.230 | 54.003 |
| AB/BA | PBE | 1.712 | 103.786 | 103.918 | 133.009 |
|  | HSE06 | 2.299 | 92.425 | 92.774 | 122.444 |

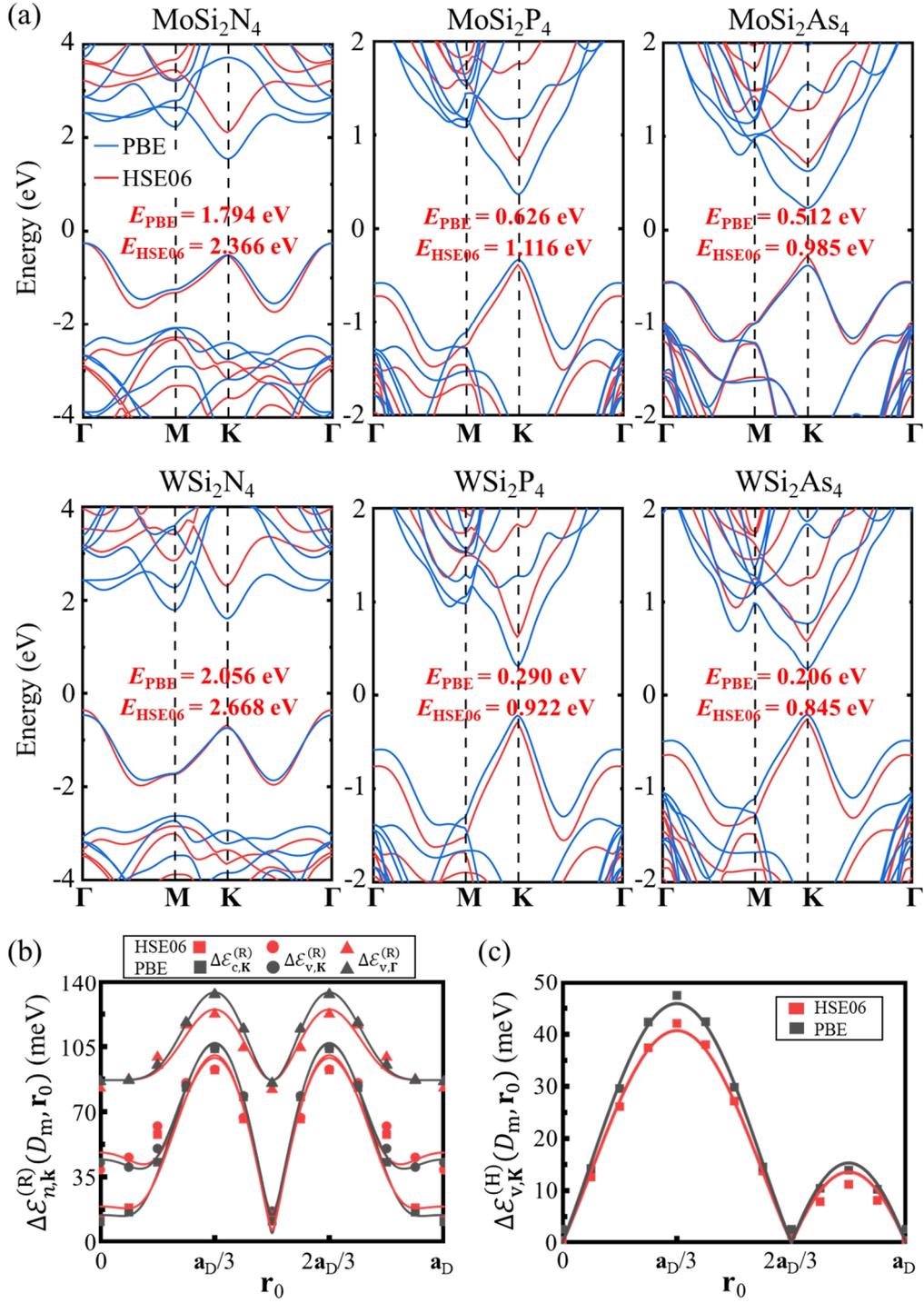

Fig. A1 (a) Comparisons between band structures of monolayer $M\text{Si}_2X_4$ calculated using PBE (blue curves) and HSE06 (red curves) functionals. SOC is not included. (b) Comparisons between energy splittings calculated using PBE (black symbols) and HSE06 (red symbols) functionals, for R-type homobilayer $\text{MoSi}_2\text{N}_4$. Square symbols correspond to $\Delta\varepsilon_{c,\mathbf{K}}^{(R)}(D_m, \mathbf{r}_0)$, circle symbols correspond to $\Delta\varepsilon_{v,\mathbf{K}}^{(R)}(D_m, \mathbf{r}_0)$, and triangle symbols correspond to $\Delta\varepsilon_{v,\mathbf{\Gamma}}^{(R)}(D_m, \mathbf{r}_0)$. The interlayer distance is set as the minimum equilibrium value $D_m$. Solid curves are fitting curves using by Eq. (4-8). (c) The energy splitting $\Delta\varepsilon_{v,\mathbf{K}}^{(H)}(D_m, \mathbf{r}_0)$ between the two valence sub-bands at $\mathbf{K}$ under $D_m$ for H-type homobilayer $\text{MoSi}_2\text{N}_4$.